%% file: main.tex
\documentclass[referee]{gji}

\usepackage{timet,color}
\usepackage[urlcolor=blue,citecolor=black,linkcolor=black]{hyperref}
\usepackage[utf8]{inputenc} % allow utf-8 input
\usepackage[T1]{fontenc}    % use 8-bit T1 fonts
\usepackage{hyperref}       % hyperlinks
\usepackage{url}            % simple URL typesetting
\usepackage{booktabs}       % professional-quality tables
\usepackage{amsfonts}       % blackboard math symbols
\usepackage{nicefrac}       % compact symbols for 1/2, etc.
\usepackage{microtype}      % microtypography
\usepackage{xcolor}         % colors
\usepackage{amsmath, graphicx}

\title{
High-resolution eikonal imaging and uncertainty quantification of the Kilauea caldera
}

\author[AF Gao et al.]
  {\Large{Angela F. Gao$^1$, John D. Wilding$^2$, Ettore Biondi$^2$, Katherine L. Bouman$^1$, 
  Zachary E. Ross$^2$} \\
  $^1$ Computing and Mathematical Sciences, California Institute of Technology, Pasadena, California, USA\\
  $^2$ Seismological Laboratory, California Institute of Technology, Pasadena, California, USA
  }
\date{}
\begin{document}

\label{firstpage}

\maketitle

\begin{summary}
% \begin{abstract}
Images of the Earth’s interior can provide us with insight into the underlying properties of the Earth, such as how seismic activity might emerge and the interplay between seismic and volcanic activity. Understanding these systems requires reliable high-resolution images to understand mechanisms and estimate physical quantities. However, reliable images are often difficult to obtain due to the non-linear nature of seismic wave propagation and the ill-posedness of the related inverse problem. Reconstructions rely on good initial estimates as well as hand-crafted priors, which can ultimately bias solutions. In our work, we present a 3D reconstruction of Kilauea's magmatic system at a previously unattained resolution. Our eikonal tomography procedure improves upon prior imaging results of Kilauea through increased resolution and per-pixel uncertainties estimated through variational inference. In particular, solving eikonal imaging using variational inference with stochastic gradient descent enables stable inversion and uncertainty quantification in the absence of strong prior knowledge of the velocity structure. Our work makes two key contributions: developing a stochastic eikonal tomography scheme with uncertainty quantification and illuminating the structure and melt quantity of the magmatic system that underlies Kilauea.
% \fix{Our work presents a new tomography scheme while also shedding light on the structure of the magmatic system and the quantity of melt.}
\end{summary}

\begin{keywords}
seismic tomography, 
volcano seismology, 
Bayesian inference, 
statistical methods, 
tomography
\end{keywords}

% Images of the Earth’s interior can provide us with insight into the underlying properties of the Earth, such as how seismic activity might emerge and the interplay between seismic and volcanic activity. 
% However, reliable high-resolution images are often difficult to obtain due to the ill-posed and non-linear nature of seismic wave propagation. 
% Reconstructions rely on good initial estimates as well as hand-crafted priors, which can ultimately bias solutions. 
% In our work, we present a 3D reconstruction of Kilauea's magmatic system at an unprecedented resolution. 
% We do this through a Bayesian approach to eikonal tomography, solving for the entire posterior distribution using variational inference. 
% Our approach provides us with the uncertainty of our reconstruction, while also shedding light on the quantity of melt parts of the magmatic system.

% \end{abstract}

\input{sections/sec_intro}

\input{sections/sec_approach}
\input{sections/sec_results}
\input{sections/sec_conclusion}
\section*{Acknowledgments}
AFG and KLB received support from Amazon AI4Science Partnership Discovery Grant, NSF award 2048237, and a Sensing to Intelligence (S2I) Award. JDW and ZER received support from NSF award EAR-2239666.

\section*{Data Availability Statement}
We will release our reconstructed $V_P$ and $V_S$ models, code, and the travel time data we used to generate them.

\bibliographystyle{plain}
\bibliography{references.bib}

\newpage

\input{sections/sec_supp}

%%%%%%%%%%%%%%%%%%%%%%%%%%%%%%%%%%%%%%%%%%%%%%%%%%%%%%%%%%%%

\end{document}

%% file: sections/sec_intro.tex
\section{Introduction}

% - What is seismic imaging and why is it useful
% - What are the challenges with TTT (ill posedness, localization) 
% - discuss localization and illposedness
% - why kilauea
% - the utility of TTT for kilauea
% - prior tomography and what they have shown
 
Kilauea volcano in Hawai`i is one of the most extensively monitored volcanoes on Earth. Since the onset of the decades-long Pu`u `O`o eruption in 1983, seismological, geodetic, and gravity investigations into the sustained eruptive activity at Kilauea have provided a rich understanding of the location of the volcano’s near-surface melt reservoirs, rift zone structure, and dike pathways and dynamics \cite{got_new_2003,lengline_tracking_2021,baker_topdown_2012,johnson_shallow_2010}. As a result, pre-eruptive patterns of deformation and seismicity originating within the upper kilometers of the subsurface can often be interpreted to assess the volcano’s immediate eruptive potential \cite{montgomerybrown_geodetic_2010,anderson_magma_2019,neal_2018_2019,klein_seismicity_1987}.

Our understanding of Kilauea's longer-term eruption cycle behavior, however, may be limited by our understanding of the lower-crustal and mantle structures that supply magma to the volcano. Kilauea is supplied with magma from a zone of melting at 80-100 km depth proximal to the Hawaiian hotspot \cite{wright_deep_2006}. Recent work has suggested that structures within Kilauea’s mantle plumbing system can have an important influence on short-term fluctuations in magma supply rate to Kilauea, which may in turn influence its eruptive behavior on timescales of weeks to years \cite{wilding_insights_2024}. During an observed “magma surge” episode in 2003-2007, the rate of magma supply to Kilauea increased by a factor of roughly two \cite{poland_mantle-driven_2012}; other work has suggested that such surges may be enabled by feedback loops between tectonic and magmatic processes within mantle magma-bearing structures \cite{wilding_insights_2024}. Furthermore, temporal correlations between eruptions at Kilauea and earthquake swarms at >30 km depth have been interpreted to evidence an “open” fluid connection between the surface and the mantle through which pressure gradients can rapidly propagate \cite{wilding_magmatic_2023}. While density anomalies constrained by gravity modeling \cite{denlinger_density_2024} and concentrations of long-period seismicity \cite{wright_deep_2006,battaglia_location_2003} suggest the existence of a columnar magma conduit extending downwards from Kilauea’s shallow reservoirs, the extent, geometry, and compositional melt fraction of this structure is still unknown. Improving our understanding of the mid-to-lower crustal structures that supply magma to Kilauea would allow us to better understand the features that enable short-term magma surges to the volcano, as well as the correlations between surface unrest and mantle seismicity.
% Improving our understanding of the sub-summit structures that supply magma to Kilauea would allow us to better understand the features that enable short-term magma surges to the volcano, as well as the correlations between surface unrest and mantle seismicity.

% passive [] and active [] travel-time tomography (TTT) studies. \fix{The earth's subsurface is studied through imaging techniques that solve an ill-posed inverse problem using travel time measurements from earthquakes to receivers at the surface. The resulting  
% \fix{The earth's subsurface is studied through imaging techniques using the time of flight from an earthquake to the surface. Katie comment: also a chance to introduce velocity. this paragraph should be where the CS person is like okay I'm getting the basic problem. I know the goal of the method, inputs and outputs, and some lingo to get through the rest of the paper.}
Numerous seismic tomography models have been developed to study subsurface structure at Kilauea \cite{ellsworth_three-dimensional_1977,thurber_seismic_1984,okubo_imaging_1997,hansen_seismic_2004,park_comparative_2007,lin2014three} but there is still significant uncertainty about the overall magma storage beneath the Hawai`i volcanic system. Prior models typically resolve features along Kilauea's rift zones, corresponding to basaltic dikes, as well as anomalies at depths of 1-12 km beneath the caldera that are usually interpreted as corresponding to the deposition of ultramafic cumulates and a magma column beneath the summit \cite{hill_geophysical_1987,ryan_mechanics_1988}. As the amount of high-quality first-arrival pick data collected at Kilauea has grown, the resolution of resolved models has steadily increased (e.g., \cite{lin2014three}). However, more detailed conclusions from such models are often hindered by challenges associated with the tomographic inverse problem.
% many inverted models are limited by the dampening needed for simultaneous source-localization and inversion involved in real world passive seismology. 

It is well known that seismic tomography is ill-posed and non-unique \cite{rawlinson_seismic_2014}, and these properties result from several factors including the sparsity of noisy observations and the non-linearity of the forward problem. Consequently, inversion results are sensitive to the initial values and rely on strong regularization (e.g., smoothing)\cite{rawlinson_seismic_2014}. Seismic imaging studies typically utilize dampening as a form of regularization, which relies on the assumption that the initial 1-D velocity model captures the first-order structure of the underlying region of interest (e.g., \cite{thurber1999simul}). However, an incomplete or inaccurate initialization can cause the inversion to be trapped in local minima due to the non-linearity of the optimization problem; thus, solving for a single reconstruction may provide a limited or biased view of the structures that are actually constrained by the available travel-time data \cite{fichtner_hamiltonian_2019}.

The uncharacterized uncertainty of the velocity, which results from the non-linear and non-unique tomography problem, limits our ability to draw robust inferences about the geometry and melt fraction of melt storage regions. For example, existing tomographic models of Kilauea generally resolve a single, undifferentiated velocity anomaly that is interpreted as a melt storage region below the caldera surface, although evidence from geodetic studies suggests the presence of multiple, distinct chambers \cite{poland_mantle-driven_2012,baker_topdown_2012}. Without quantitative information about model uncertainty, it is difficult to assess these models’ ability to differentiate between a single spatially distributed melt storage region or multiple neighboring chambers. Furthermore, placing confidence bounds on melt fraction values calculated from velocity values within imaged melt storage regions (e.g., \cite{maguire_magma_2022}) requires an estimate of uncertainty in the model parameters, which is not available in typical tomographic workflows.

In this study, we develop an eikonal tomography procedure and apply it to produce a high-resolution tomographic image of the crust beneath the Kilauea summit. Our technique leverages advances in statistical inference to produce first-order uncertainty estimates through variational inference on recovered tomographic models. In particular, we use stochastic optimization through Monte Carlo sampling, which enables stable inversion in settings with weak or no priors. We validate our technique on realistic test cases and then apply our technique to travel-time data recorded at Kilauea volcano and identify a distinctly high-$V_P$/$V_S$ region beneath the summit that we interpret as the magma-bearing column connecting the surface reservoirs to magma supply from the mantle. We take advantage of our derived uncertainties to estimate the size of this column and constrain a melt fraction of roughly 15\% within this column. Our results suggest the existence of persistent melt structures connecting Kilauea to the mantle that may enable feedback between the surface and deeper parts of the magma plumbing system.

%% file: sections/sec_approach.tex
\section{Background}\label{sec:approach}

\begin{figure}
    \centering
    \includegraphics[width=.5\textwidth]{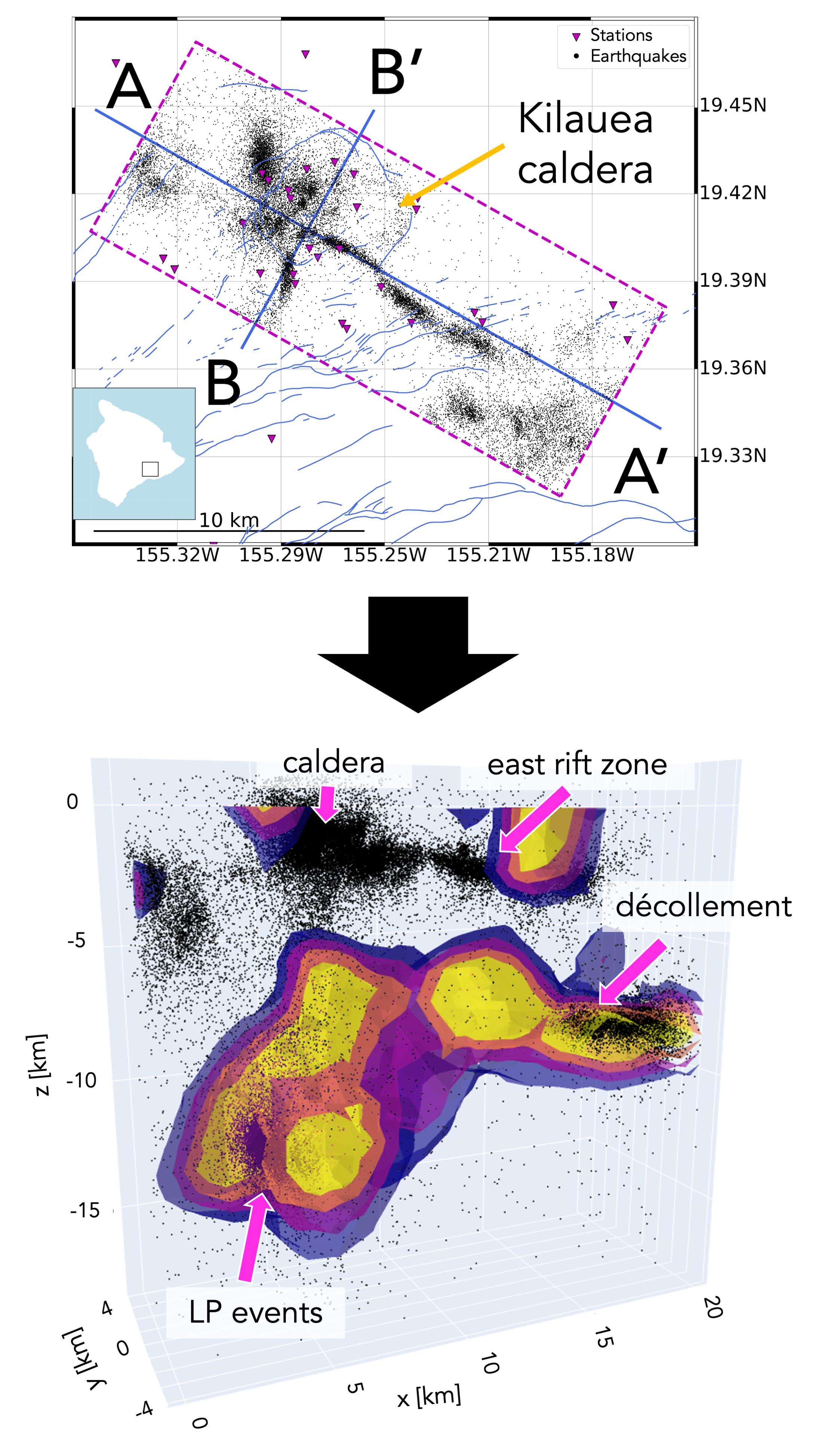}
    \caption{\textbf{Source-receiver geometry and recovered 3D magmatic system from a subdomain of the Island of Hawai'i.} For our real data study and realistic-geometry synthetic tests, we consider the sources and receivers in a small region of the Island of Hawai'i that includes the Kilauea caldera. We use a total of 18 receivers and 45,310 earthquake events. These events were located using the method described in \protect\cite{wilding2023magmatic}, but with modifications discussed in Sec.~\ref{sec:catalog_details}. We visualize the cross sections A-A' and B-B' used for later figures. The bottom figure depicts our reconstruction of the 3D magmatic system that underlies Kilauea. }
    \label{fig:geo}
\end{figure}
\subsection{Travel time tomography (TTT)}
Traditionally, seismic tomography using travel times has been solved using ray-based tomography techniques \cite{thurber1999simul,zhang2003tomodd,zhang2006development}. These methods aim to solve for a heterogeneous velocity model through modeling the ray trajectory, which are curved by the heterogeneous velocity model. However, continuously curved ray tracing is expensive, so in practice, ray-based tomography often uses pseudo-ray bending \cite{um1987fast}, which relaxes the curved ray tracing problem into a piecewise linear approximation to improve efficiency. Ray-theory-based tomography techniques \cite{zhang2003tomodd,thurber1999simul} are often very efficient, but they are only an approximation of continuous ray tracing, which is a weak approximation of the wave equation.

Imaging using the eikonal equation has recently become possible due to improvements in computation. The eikonal equation, which is an infinite frequency approximation of the wave equation, underlies the arrival times of earthquakes and governs the travel time along the shortest path between two points \cite{rawlinson2005fast}. Let the travel time for a seismic wave be defined as $T_V$. The eikonal equation relates the travel time to velocity and is defined by 
\begin{equation}
    ||\nabla_{\textbf{x}_r} T_V(\textbf{x}_s, \textbf{x}_r)||^2 = \left(\frac{1}{V(\textbf{x}_r)}\right)^2
\end{equation}
where $T_V(\textbf{x}_s, \textbf{x}_r) \in C^1(\mathbb{R}^{2d})$ is the travel time field between source location $\textbf{x}_s \in \mathbb{R}^d$ and receiver location $\textbf{x}_r\in \mathbb{R}^d$ induced by the velocity model $V$ and $V(\textbf{x}_r)$ is the velocity at a location $\textbf{x}_r$. In other words, the eikonal equation states that the velocity $V(\textbf{x}_r)$ at a location $\textbf{x}_r$ is inversely proportional to the norm of the gradient of the travel time field $T_V(\textbf{x}_s, \textbf{x}_r)$ with respect to that location $\textbf{x}_r$. For an earthquake's P and S waves, the associated arrival times are induced by separate velocity models, which we denote as $V_P$ and $V_S$ respectively.

% Travel time measurements are governed by the Eikonal equation, which relates the arrival time to slowness, which is inversely proportional to the velocity of the medium. 
% \begin{equation}
%     |\nabla_r T(x_s, x_r)|^2 = \left(\frac{1}{V(x_r)}\right)^2
% \end{equation}

It has become relatively common in recent years to solve seismic imaging problems with Bayesian inference \cite{zhao2022bayesian,gao2021deepgem}.  Here, prior knowledge about the velocity can be incorporated when available to help stabilize the inverse problem like in \cite{muir2020geometric}. In the simplest setting, one can solve for the minimizer of the \textit{maximum a posteriori} (MAP) objective. In particular, the MAP approach for TTT aims to find the velocity model $V$ (e.g., $V_P$ or $V_S$) that maximizes the following objective given travel time measurements $y$ (e.g., P wave or S wave travel times)
\begin{equation}
    \hat{V} = \arg\max_{V} \underbrace{\log p(y|V)}_{\text{log data likelihood}} + \underbrace{\log p(V)}_{\text{log prior}}. \label{eq:map_basic}
\end{equation}
Often, the travel time measurement from sources $\{\textbf{x}_{s_i}\}_{i\in \mathcal{S}}$ and receivers $\{\textbf{x}_{r_j}\}_{j \in \mathcal{R}}$ induced by a velocity model $V$ are assumed to be of the form 
\begin{equation}
\{y_{ij} = T_V(\textbf{x}_{s_i}, \textbf{x}_{r_j})+ \varepsilon_{ij}\}_{i \in \mathcal{S}, j \in \mathcal{R}}    \label{eq:measurement}
\end{equation}
where $\mathcal{S}$ and $\mathcal{R}$ represent the set of sources and receivers respectively. The measurement noise $\varepsilon_{ij}$ is commonly assumed to follow a Gaussian distribution $\varepsilon_{ij} \sim \mathcal{N}(0, \sigma_{ij})$. This measurement noise model produces the data likelihood model  $y_{ij} \sim \mathcal{N}(T_V(\textbf{x}_{s_i}, \textbf{x}_{r_j}), \sigma_{ij})$, which results in the $\log$ data likelihood, $\log p(y|V)$, taking the form of a weighed sum of squared error (SSE):

\begin{equation}
    \hat{V} = \arg\max_{V} \underbrace{\sum_{ij} \frac{1}{\sigma_{ij}^2} (y_{ij}-T_V(\textbf{x}_{s_i}, \textbf{x}_{r_j}))^2}_{\text{log data likelihood}} + \underbrace{\log p(V)}_{\text{log prior}}. \label{eq:map}
\end{equation} 

% Sometimes, there are additional constraints that cannot be encoded in a true prior probability as in Eq.~\ref{eq:map}. In those cases, Regularized Maximum Likelihood (RML) is used where regularizes are used in place of priors. Regularizers $\mathcal{R}(V)$ replace $\log p(V)$ to solve the following objective: 
% \begin{equation}
%     \hat{V} = \arg\max_{V} \log p(y|V) + \mathcal{R}(V).
% \end{equation}

Eikonal-based travel time calculations can be advantageous to those derived from traditional ray-tracing due to their greater stability in the presence of strong velocity heterogeneity \cite{lin2009eikonal,de2011ambient,tong2021adjoint,biondi2023upper,gao2021deepgem}.  Historically, however, eikonal TTT has been less popular; this may be due to various factors including particular challenges of solving the eikonal equation itself \cite{rawlinson2005fast,fomel2009fast}, added computation time and memory, and the limited availability of strategies for solving these inverse problems. Recently \cite{biondi2023upper} developed an adjoint-state approach for eikonal TTT that requires relatively low memory overhead, enabling the approach to scale to millions of source-receiver pairs. In our approach, we utilize this adjoint operator  to update the inferred velocity model.

% Unlike ray-tracing based inversion, eikonal-based methods are more robust to heterogeneous velocity models. However, the eikonal equation does not allow for wave effects from indirect arrivals. Since we are only considering imaging using direct arrivals, 

\subsection{Variational inference in seismic imaging}

Posterior estimation is a useful tool in ill-posed imaging problems because it provides estimates of uncertainty for image reconstruction. Two popular classes of posterior estimation methods are Markov Chain Monte Carlo (MCMC) \cite{betancourt2015hamiltonian} and variational inference (VI) \cite{blei2017variational}. Although MCMC is exact and asymptotically converges to the true posterior, it is computationally demanding to reach the true solution, which is hard to validate. Moreover, it scales poorly to high dimensional posterior distributions and can be prohibitively expensive computationally even for 2D imaging problems\cite{sun2021deep}. On the other hand, VI is much more efficient at estimating the posterior by constraining estimation over a tractable class of parameterized candidate distributions, but approximates the distribution through this tractable candidate set, making it less exact. In our work, we focus on using VI for uncertainty quantification due to the high dimensional 3D imaging case that we consider.

VI poses the Bayesian posterior inference problem as one of optimization, in which the the target distribution is approximated with a parameterized candidate distribution \cite{sun2021deep}. This candidate distribution is the one that best fits the target distribution and is a member of a pre-specified class of distributions, referred to as the \textit{variational class}. It is found through optimization by minimizing a loss (e.g., Kullback Leibler divergence \cite{kullback1951information}) between distributions through $\arg\min_{\theta} D_{\text{KL}}(q_{\theta}||p)$. The choice of variational class is extremely important to well-estimate the posterior, but having more complex variational classes can make the variational inference problem less tractable. There is a trade-off that must be considered between the expressivity of the variational distribution and the efficiency of the optimization problem.

Prior works in seismic imaging performing uncertainty quantification for TTT using VI include \cite{zhang2020seismic,zhao2022bayesian}. In particular, \cite{zhang2020seismic} looks at posterior estimation using ADVI and Stein variational gradient descent (SVGD) while \cite{zhao2022bayesian} uses normalizing flows as the variational distribution. There are other works that study uncertainty quantification for full-waveform inversion (FWI) using VI including \cite{zhang2020variational,zhang20233} which use SVGD for FWI. 
% Although \cite{zhang20233} is an example showing UQ for 3D FWI, it has only been demonstrated with one inversion requiring that all choices of priors and hyperparameters were correct.
 
\subsection{Hypocenter locations} \label{sec:SSSTs}
Earthquake hypocenters are almost always determined by solving an inverse problem and thus subject to error, which ultimately can impact the resulting tomography. However, accurate locations depends on accurate velocity models, creating significant trade-off between the two quantities~\cite{zhang2006development}. For this reason, iterative methods that alternate between localization and tomography are popular to prevent bias from poor hypocenter locations \cite{biondi2023upper,thurber1999simul,zhang2003tomodd}. However, accurately estimating the joint uncertainty in the retrieved hypocentral locations and tomographic models is challenging in this iterative framework because of non-convexity.

One way to circumvent the need for iterative methods is to better localize earthquakes initially by implicitly capturing the unmodeled 3D structure without systematically biasing the locations. A popular technique to do this is using source specific station correction terms (SSSTs) when locating events \cite{richards2000earthquake,lin2005tests}. Given the origin times (i.e. the time of earthquake rupture), which are solved for concurrently, the hypocenter locations corrected with SSSTs are found by solving the following objective:
% \begin{equation}
    % \hat{x}_{s_i} = \arg\min_{x_{s_i}, \{\tau_{ij}\}_{j \in \mathcal{R}}} \sum_{j} \gamma_{ij}||(T(V, x_{s_i}, x_{r_j})+ \tau_{ij}) - y_{ij}|| \label{eq:sssts}
% \end{equation}
\begin{align}
    \{\hat{\textbf{x}}_{s_i}\}_{i \in \mathcal{S}} &= \arg\min_{\substack{\{\textbf{x}_{s_i}\}_{i \in \mathcal{S}} \\ \{\tau_{ij}\}_{(i,j) \in \mathcal{S} \times \mathcal{R}}}} \sum_{i,j} \gamma_{ij}||(T_V(\textbf{x}_{s_i}, \textbf{x}_{r_j})- \tau_{ij}) - y_{ij}|| \label{eq:sssts}\\
     \qquad &\text{s.t.} \quad \tau_{ij} = \text{Median}(\{\delta_{kj}\}_{kj}) \quad \text{where} \nonumber \\
    &  \qquad \delta_{kj} = |T_V(\textbf{x}_{s_k}, \textbf{x}_{r_j}) - y_{kj})|\quad \text{for} \quad \textbf{x}_{s_k} \in\mathcal{B}_r(\textbf{x}_{s_i} )\nonumber
\end{align}
where the SSST value, $\tau_{ij}$, corresponding to each travel time measurement $y_{ij}$ is found through iteratively shrinking the neighborhood $\mathcal{B}_r(\textbf{x}_{s_i})$, a ball of radius $r$ and $\gamma_{ij}$ weighs each neighbor based on the quality of the measurement. Essentially, this implicitly incorporates the unmodeled 3D structure through estimating a local correction $\tau_{ij}$ and finds the optimal location based on this implicit structure. 

% This optimization procedure is typically iterated a few times until the box size is sufficiently small. Unlike \cite{wilding2023magmatic}, we choose not to utilize GrowClust \cite{trugman2017growclust} to further refine the locations because it can cause significant systematic bias to the hypocenter locations.
% \fix{$x_s$ is $x, y, z, t$ with $t$ being the origin time}

\section{Method}\label{sec:approach}
% In this paper, we demonstrate a method to perform uncertainty quantification with variational inference for high resolution Eikonal tomography with application to the Kilauea crater. 

% \begin{figure}
%     \centering
%     \includegraphics[width=.5\textwidth]{FiguresPDFs/Figures-Kilauea.png}
%     \caption{\textbf{Source receiver geometry from a subdomain of the Island of Hawai'i.} For our real data study and realistic-geometry synthetic tests, we consider the sources and receivers in a small region of the Island of Hawai'i that includes the Kilauea caldera. We use a total of 18 receivers and 45,310 earthquake events. These events were located using the method described in \protect\cite{wilding2023magmatic}, but with modifications discussed in Sec.~\ref{sec:catalog_details}. In the upper right figure, we visualize the cross sections A-A' and B-B' used for later figures. The bottom figure depicts our reconstruction of the 3D magmatic system that underlies Kilauea.}
%     \label{fig:geo}
% \end{figure}

Our variational method performs high resolution eikonal tomography without the need to relocate hypocenters during the inversion procedure. 
% Additionaly, we compare various design choices and variational distributions to better understand the trade-offs made between computational cost and accuracy.
Our approach aims to solve for the minimizer of the Kullback-Leibler (KL) divergence, a commonly used objective for VI, which aims to find the optimal recovered posterior distribution $q_{\theta}(V)$. In particular, our goal is to find the optimal set of parameters $\theta$ for our inferred posterior distribution $q_{\theta}(V)$ which belongs to some predefined variational distribution $\mathbb{P}$. The optimization problem we aim to solve is
% \textcolor{red}{right now x is in place of V but notation will probably be changed elsewhere}
\begin{align}
 \theta^* &=\arg\min_{\theta\in \mathbb{P}}D_{\text{KL}} (q_{\theta}(V) || p(V|y))\nonumber\\
    &= \arg\min_{\theta\in \mathbb{P}}E_{V \sim q_{\theta(V)}} [\log q_{\theta}(V) - \log p(V|y)]\nonumber\\ 
    &= \arg\min_{\theta\in \mathbb{P}}E_{V \sim q_{\theta(V)}} [\log q_{\theta}(V) - \log p(y|V) - \log p(V) \nonumber\\
    & \qquad \qquad \qquad \qquad \qquad + \log p(y)]\nonumber\\ 
    &= \arg\min_{\theta\in \mathbb{P}}E_{V \sim q_{\theta(V)}} [\underbrace{\log q_{\theta}(V)}_{\text{neg log entropy}} - \underbrace{\log p(y|V)}_{\text{log data likelihood}} - \underbrace{\log p(V)}_{\text{log prior}}]. \label{eq:DPI}
\end{align}
The $\log$ data likelihood takes on different forms depending on how the added noise is modeled. For instance, a very common assumption is that the measurements have i.i.d. Gaussian additive noise $\varepsilon_{ij} \sim \mathcal{N}(0, \sigma)$ where the $\log$ data likelihood is $\log p(y_{ij}|V) = \frac{1}{\sigma^2}||T_V(x_{s_i}, x_{r_j}) - y_{ij}||_2$ and the measurements are assumed to be of the form $y_{ij} \sim T_V(x_{s_i}, x_{r_j}) + \varepsilon_{ij}$.
Another setting that we consider is when the noise is i.i.d. Laplacian additive noise $\varepsilon_{ij} \sim Laplace(0, b)$ where the $\log$ data likelihood is $\log p(y_{ij}|V) = \frac{1}{b} ||T_V(x_{s_i}, x_{r_j}) - y_{ij}||_1 $ and the measurements are of the form $y_{ij} \sim T_V(x_{s_i}, x_{r_j}) + \varepsilon_{ij}$. In this work, we consider the setting where the prior, $\log p(V)$, is implicit and consequently ignored in Eq.~\ref{eq:DPI}. In particular, we assume an implicit prior through gradient smoothing that is described in Sec.~\ref{sec:gradsmoo} and set $\log p(V)$ in \ref{eq:DPI} to be $ 0$.

For our variational family, we use a multivariate $\log$ normal distribution of the form $V \sim \operatorname{Lognormal}(\mu, \sigma I)$ where $\mu, \sigma \in \mathbb{R}^{n_x \times n_y \times n_z}$ for $V \in \mathbb{R}^{n_x \times n_y \times n_z}$. 
The $\log$ normal variational distribution ensures that the velocity is non-negative. Our inferred posterior distribution $q_{\theta}(V)$ is the distribution that minimizes the negative $\log$ data likelihood ($-\log p(y|V)$), minimizes the negative $\log$ prior ($-\log p(V)$), and maximizes the entropy ($- \log q_{\theta}(V)$) of the distribution, which encourages diversity of the samples of the inferred posterior.

We estimate the expectation in Eq.~\ref{eq:DPI} using Monte Carlo by averaging over samples from the inferred posterior distribution $q_{\theta}(V)$ giving
\begin{align}
    \theta^* \approx \arg\min_{\theta \in \mathbb{P}} \frac{1}{N}&\sum_{i=1}^N [\log q_{\theta}(V^{(i)}) - \log p(y|V^{(i)})]\nonumber \\
    &\quad \text{for} \quad \log p(V^{(i)})=0. \label{eq:DPI_MC}
\end{align}
We solve the optimization problem in Eq.~\ref{eq:DPI_MC} using gradient descent and initialize the velocity model to be the 1D model used for source localization in \cite{wilding2023magmatic}. The uncertainty is computed through sampling the distribution and computing per-pixel standard deviations. Unlike second order methods like BFGS, gradient based methods allow for stochastic gradients, which are induced from the Monte Carlo sampling procedure. We use Adam \cite{kingma2014adam} as the optimizer.

\subsection{Gradient smoothing} \label{sec:gradsmoo} 
Many inverse problems are ill-posed and require the usage of priors to prevent overfitting to suboptimal solutions. In seismology, tomography methods typically employ Laplacian regularization to penalize roughness and avoid overfitting to noise. 
% Explicit regularization of this form can yield inversion results that are biased by the initialization.

An alternative regularization approach, designed for iterative inversion algorithms based on gradient descent, smooths the gradient of $\log p(y|V)$ from Eq.~\ref{eq:DPI} spatially to avoid overfitting. We adopt this latter approach in our study and spatially smooth the gradient of $\log p(y|V)$  annealing the degree of smoothing over the course of the optimization. We implement the gradient smoothing with a 3D isotropic Gaussian blur through the following parameter update rule,
\begin{equation}
    V^{(k+1)} = V^{(k)} -\eta G(r^{(k)})\frac{\delta{\mathcal{L}}}{\delta V } \label{eq:grad_update}
\end{equation}
where $\eta$ is the learning rate, $G$ is a Gaussian spatial smoothing operator, $r^{(k)}$ is the radius of isotropic Gaussian blur that monotonically decreases over epochs $k$, and $\mathcal{L}$ is the objective as defined in Eq.~\ref{eq:map} or Eq.~\ref{eq:DPI_MC}. Annealing the smoothing operator improves the optimization landscape by preventing overfitting early on in the optimization. This is equivalent to multi-scale inversion approaches under certain assumptions, and further analysis is included in the appendix.

We consider various ways to define the smoothing radius $r^{(k)}$ as a function of epoch $k$, each of which begins with relatively strong gradient smoothing and ends with relatively weak smoothing. In particular, we consider a piecewise function for the radius $r^{(k)}$ of the isotropic Gaussian blur that initially decays exponentially and then becomes constant after $k_{\text{stop}}$ epochs: 
\begin{equation}
    r^{(k)} = 
\begin{cases}
    5e^{-3k/k_{\text{max}}} & k < k_{\text{stop}} \\
    5e^{-3k_{\text{stop}}/k_{\text{max}}} & k \geq k_{\text{stop}} \label{eq:gradsmoo}
\end{cases}.
\end{equation}
In our synthetic tests, we consider various values of $k$, $k_{\text{max}}$, and $k_{\text{stop}}$.  For our imaging results with real data, we use $k=250$, $k_{\text{max}} = 500$, and $k_{\text{stop}} = 250$, as in explained in Sec.~\ref{sec:gradsmoo_synth}.

%% file: sections/sec_results.tex
\section{Synthetic imaging}\label{sec:synthetic_results}

\label{sec:experiments}

% \subsection{3D synthetic anomaly tests}

% We perform synthetic tomography tests modelled after a region of the big island of Hawaii using a subset of real stations and real event locations from \cite{wilding2023magmatic}. We look at a domain from \textcolor{red}{add lat lon} using \textcolor{red}{1000 events and 8 receivers} at a resolution of .5 km. 

% \paragraph{Changing initialization} We perform various anomaly and hyperparameter tests to understand the sensitivity of our measurements. We consider various initial conditions from \cite{}.

In this section, we show results from synthetic tests to understand the intrinsic resolution of the imaging setup, evaluate the quality of the uncertainty quantification, and evaluate the effect of various algorithmic choices on the resulting tomography. 
% We apply our method to perform uncertainty quantification for eikonal tomography of a region of Hawai'i underlying the volcano Kilauea. 

For our synthetic tests, we take the same catalog of 45,310 events as described in Sec.~\ref{sec:catalog_details} and generate synthetic arrival times between all 18 receivers for a total of 815,580 P and S arrival time measurements each.
We also assume the same measurement noise as in Sec.~\ref{sec:catalog_details} where the observed travel time between source $x_{s_i}$ and receiver $x_{r_j}$ is $y_{ij} = T_V(x_{s_i}, x_{r_j}) + \eta_{ij}$ for $\eta_{ij} \sim Laplace(0, b)$. The size of the domain used is approximately 25 km $\times$ 10 km $\times$ 20 km. Our reconstructions are performed on a uniform grid of 1 km$^3$ in size.

To understand the sensitivity of reconstruction to various algorithmic choices, we perform synthetic tests with the actual source-receiver geometry to inform hyperparameter choices when imaging with the real dataset. These main design choices that we consider are 1. gradient smoothing hyperparameters and 2. assumed measurement noise model. 
% Unlike prior tomography papers \cite{zhang2003tomodd}[cite], we do not consider dampening as a prior and instead use gradient smoothing as our regularization.
We also perform various resolution tests through checkerboard and anomaly tests to validate which regions we believe to have sufficient sensitivity.

% In \ref{fig:synth_mismatch_prior}, we compare the affect of the strength of prior and mismatch between the assumed noise model and defined log data-likelihood. We assume the prior to be a Gaussian distribution centered at the initialization. \textcolor{red}{todo placeholder:  As expected, increasing the weight of the prior shifts the result to fit the prior mean, effectively dampening the solution. Decreasing the weight too much can result in overfitting to the measurement noise. Underestimating the noise causes overfitting while overestimating the noise increasing the prior strength. Using a Gaussian vs. Laplacian noise model that are well matched does not affect the reconstruction substantially. }
\subsection{Hyperparameter tests}

% There are algorithmic choices that we make to perform such an inversion: 1. gradient smoothing, 2. choice of prior, 3. assumed noise model. 
\subparagraph{Gradient smoothing}\label{sec:gradsmoo_synth} 
In Fig.~\ref{fig:param_UQ}, we consider various functions of decreasing the radius $r^{(k)}$ of isotropic Gaussian blur as described in Eq.~\ref{eq:gradsmoo}.
Although in most cases, the data fit is quite similar at the final point of the optimization, they have varying degrees of overfitting to measurement noise. We find that the result from early stopping with hyperparameters $k = 250$, $k_{\text{max}} = 500$, and $k_{\text{stop}} = 250$ seems to balance fitting the data while preventing overfitting to noise for both $V_P$ and $V_S$. 

% \paragraph{Sensitivity to priors} 
% A common tool in seismic tomography used to prevent overfitting is dampening. This is often done by enforcing a small update to the initial model, which can be equivalently viewed as a prior centered at the initial velocity model. While this is a useful tool to prevent overfitting to measurement error, we choose to instead utilize gradient smoothing as our regularization instead.

\subparagraph{Sensitivity to noise model} 
One challenge with performing Bayesian tomography using real data is that it requires a forward model specifying the noise distribution, but the true measurement noise is difficult to characterize. This is due to the complex nature of instrumentation noise and arrival time picking errors, as well as the hypocenter algorithm itself. For a rough measurement noise model characterization, we use the data residuals from the seismicity catalog for the noise. We find that this is a Laplacian distribution as shown in \ref{sec:noisemodel}. However, we could be incorrectly mischaracterizing the true noise model, so we test the effects of mischaracterized noise distributions. We do this by performing synthetic tests where we make different assumptions about the true added measurement noise model and the assumed noise model in Fig.~\ref{fig:synth_mismatch}. We consider the measurement noise $\eta_{ij}$ drawn either from Gaussian or Laplacian distribution and the assumed measurement models for the data likelihood  $\log p(y|V)$ either to be Gaussian or Laplacian. From our synthetic tests in Fig.~\ref{fig:synth_mismatch}, there are some differences in the reconstructions in terms of the size and amplitude of the recovered anomalies, but we do not find significant differences in the recovery of large features when these noise models are interchanged. We perform additional tests where we deliberately underestimate and overestimate the amount of noise, as shown in Fig.~\ref{fig:synth_noiselevel}. In the case where we underestimate the noise, we find that early stopping helps to prevent overfitting to the measurement noise. When we overestimate the noise, we find that the resulting anomalies that are reconstructed are of a slightly lower amplitude. However, in both cases, the main features are correctly localized.

\begin{figure*}
    \centering
    \includegraphics[width=.7\textwidth]{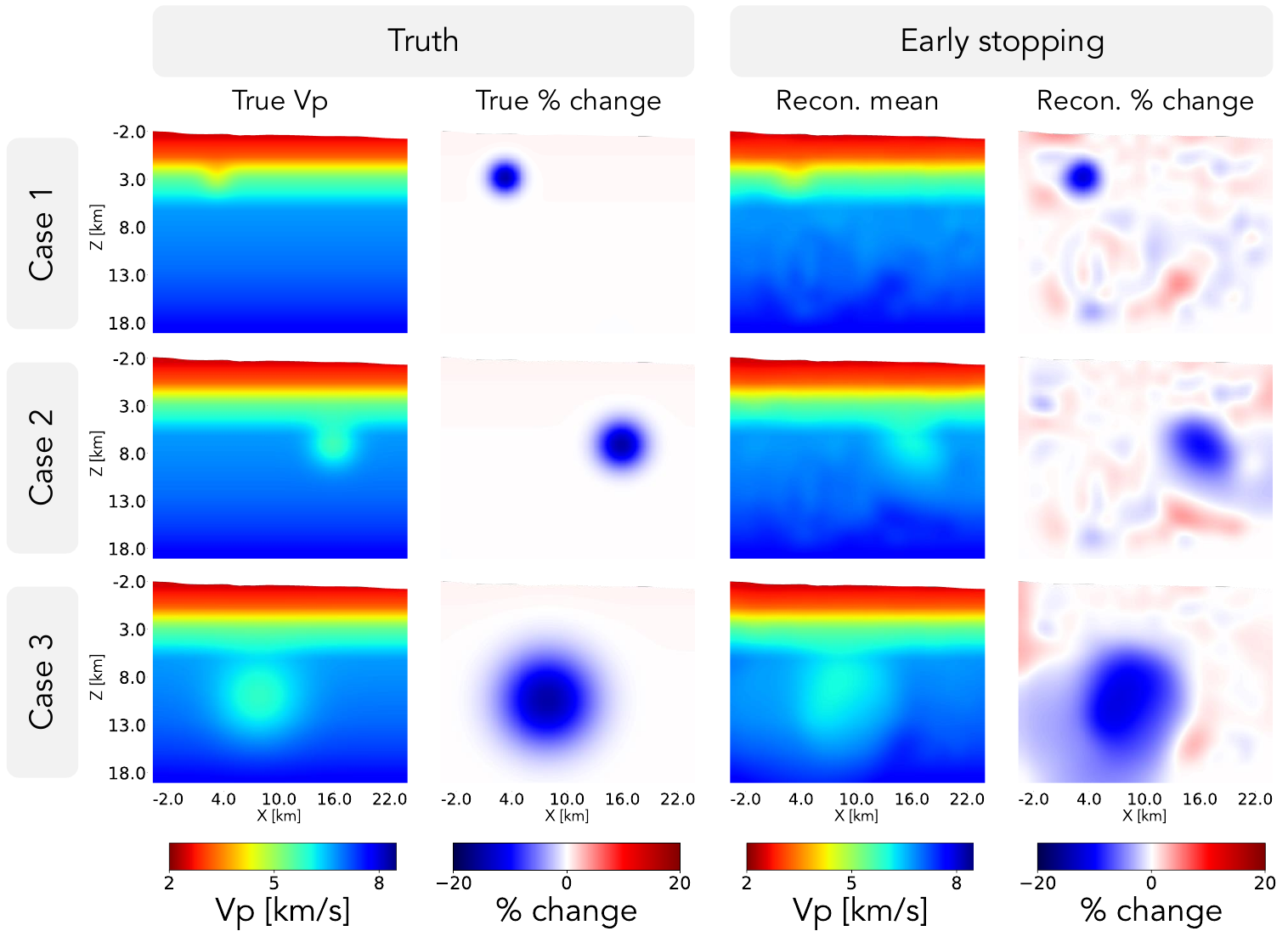}
    \caption{\textbf{Physically motivated synthetic anomaly tests.} We perform synthetic tests to explore the sensitivity of our expected source-receiver geometry to various $\sim 10\%$ slower velocity anomalies in the subsurface. For various candidate anomalies, we show the following for A-A' (shown in Fig.~\ref{fig:geo}): the true velocity model used to generate the synthetic travel times, the relative change to the initial model, and the mean of the inferred posterior and relative change to the initial model. Note, we plot these cross sections masking out the topography.  }
    \label{fig:anomaly_partial}
\end{figure*}

\begin{figure*}
    \centering
    \includegraphics[width=.9\textwidth]{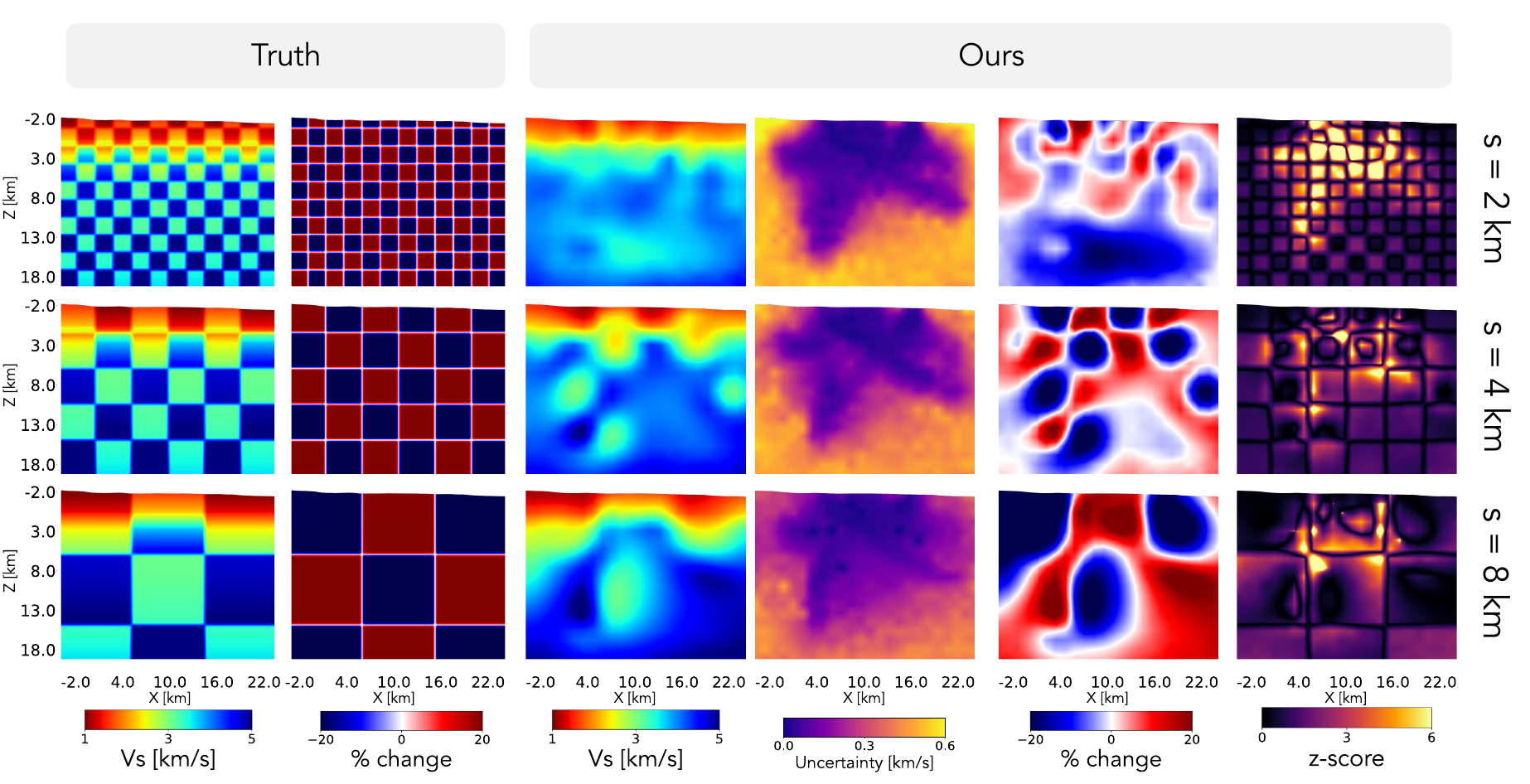}
    \caption{\textbf{Checkerboard tests}  We perform inversion on the 1D model from \protect\cite{lin2014three} modulated with various checkerboards of $25\%$ increase or decrease with a side length of $s$. These checkerboard patterns have a size of 2 km, 4 km, and 8 km.  We also show the mean and standard deviation of our reconstruction and relative change between the 1D model and our reconstruction for the A-A' cross-section. We include z-scores to quantify the accuracy of our uncertainty quantification. We are able to resolve around a resolution of 4 km in the upper half of the domain and at least a resolution of 8 km throughout the region with ray coverage. Interestingly, the estimated uncertainty for $s=4$ km is higher in low velocity regions such as the one at around 6 km depth and between 6 and 10 km in $X$, which is expected behavior. We include the full results for both $V_P$ and $V_S$ in the appendix. }
    \label{fig:check}
\end{figure*}

 % We find that the uncertainty is underestimated near the sharp interfaces of the checkerboard as well as in the high frequency regions that are below the resolvability of our method.

\subsection{Resolution tests}\label{sec:res_tests}
\subparagraph {Anomaly tests} In Figure~\ref{fig:anomaly_partial}, we perform various synthetic anomaly tests to identify how sensitive our source-receiver geometry is to the spatial and amplitude variations of potential anomalies. These anomalies are motivated by several questions of scientific interest about the subsurface plumbing structure at Kilauea. Case 1 considers small magma chambers located near the caldera, Case 2 considers medium-sized anomalies in the rift zone, and Case 3 considers deep magmatic systems underneath the Kilauea caldera. We find that our source-receiver geometry has sufficient coverage to localize and reconstruct 
Case 1, which is close to the surface where the majority of events are located. There are some artifacts in the reconstruction due to the synthetic noise added to the inverse problem. 
% \fix{The uncertainty in this posterior estimation is predominately along the ray paths that intersect with this anomaly.}
The anomaly in Case 2 is well identified, but not as well reconstructed and localized, due to the asymmetrical source-receiver geometry that results in some smearing artifacts. Again, there are artifacts due to the synthetic noise added to the inverse problem. 
% \fix{There is little uncertainty in the high velocity deep region unlike in the other two cases.} 
In Case 3, our method recovers a large deep anomaly; however, the relative shape, location, and amplitude is not exactly correct and there are some smearing artifacts. In all cases, the anomalies are reconstructed with some slight perturbations due to the source-receiver geometry. These tests give confidence in being able to recover anomalies in these regions even though the exact shape and structure is sometimes smeared.
% There is relatively high uncertainty throughout the entire region with sensitivity in Case 3. 

We also consider variations on Case 3 where the anomaly is much smaller in Fig.~\ref{fig:anomaly}. In Case 3.1, the anomaly is half the size of Case 3, and in Case 3.2, the anomaly is a fourth of the size of Case 3. In Case 3.1, the anomaly is recovered but there are more artifacts, and in Case 3.2, the recovered anomaly appears to be on the order of the measurement noise artifacts. In all of these cases, the early stopping parameters, as mentioned in Sec.~\ref{sec:gradsmoo_synth}, help prevent overfitting.

\subparagraph{Checkerboard tests} In Figure~\ref{fig:check}, we test the inherent resolution of the source-receiver geometry by using various checkerboard patterns. Using the selected hyperparameters, as mentioned in Sec.~ \ref{sec:gradsmoo_synth}, a vast majority of the region where we have earthquake-coverage is resolvable to a resolution of at least $8$ km. In the primary regions of interest, we can achieve a resolution of $4$ km. In Fig.~{\ref{fig:check_vp_vs}}, we include z-scores and find that for $s= 4$ km and $s=8$ km, the ground truth is predominantly within three to four standard deviations from the mean. Most of the high z-score regions are close to the boundaries of the checkerboard, which are sharp and hard to recover. For $s=2$, we do not trust both the recovered structure and uncertainty because of a mismatch between the blur induced by the early stopping and the resolution of the underlying model we are trying to recover. For the complete result of $V_P$ and $V_S$, see Fig.~{\ref{fig:check_vp_vs}}.

To understand the regions for which we have sensitivity (for a certain resolution size), we use both the checkerboard test and the ray coverage map to produce a resolvability index map. We borrow the resolvability index introduced in \cite{huang2015yellowstone} and used in \cite{biondi2023upper}. $R$ is defined as 
\begin{align}
    R_s & = \frac{\sum_{i \in \mathcal{D}}(\%_{\text{recon}} + \%_{\text{truth}})^2}{2\sum_{i \in \mathcal{D}}(\%_{\text{recon}}^2 + \%_{\text{truth}}^2)} \nonumber \\
   & s.t. \quad \%_{\text{recon}} = \frac{\hat{V}_c-V_0}{V_0}, \%_{\text{truth}} = \frac{V_c-V_0}{V_0}\label{eq:sensitivity}
\end{align}
where $\mathcal{D}$ is the a chosen domain size, $\hat{V}_c$ is the recovered volume with a $s$ sized checkerboard, $V_c$ is the ground truth with an $s$ sized checkerboard, and  $V_0$ is the 1D model \cite{wright2006deep}. The resolvability index $R$ ranges from $0$ to $1$ where $0$ represents perfectly opposite recovery, $1$ represents perfect recovery, and $0.5$ represents no recovery. We use $D=0.6$ as in \cite{biondi2023upper}. We define the domain $\mathcal{D}$ as slightly larger than the size of the checkerboard, and we use the checkerboard with grid size $4$ km as shown in Fig~\ref{fig:real_meas}.

\section{Imaging the Kilauea volcano}\label{sec:realdata_results}
In this section, we describe our imaging method, from raw data to final inversion result for $V_P$ and $V_S$, velocities for the P and S waves respectively. Here we describe the procedure of generating our catalog, define our measurement model, and show results from our eikonal tomography method. We find that we are able to perform reliable tomography without the need for relocation during the tomography process, unlike previous tomography approaches. By implicitly modeling the 3D structure when determining the initial hypocenters, we are able to estimate the velocity in one-shot, which allows for streamlined uncertainty quantification for the velocity model. The size of the domain used is approximately 25 km $\times$ 10 km $\times$ 20 km. Our reconstructions are performed on a uniform grid of 1 km$^3$ in size. This setup is consistent with our synthetic tests.

\subsection{Data and preprocessing procedure}\label{sec:catalog_details} 
To generate a travel-time dataset, we use a modified version of the catalog-generating methodology presented in \cite{wilding2023magmatic}. \cite{wilding2023magmatic} built a catalog for south Hawai`i spanning May 2018 to April 2022 by applying shrinking-box source-specific station terms (SSSTs) \cite{richards2000earthquake} to the P and S arrival times at the receivers, known as picks. These picks are identified by the automated phase-picking algorithm PhaseNet~\cite{zhu2019phasenet}. Inspired by recent work demonstrating that greater picking accuracy in volcanic environments can be achieved using phase-picking models trained on volcano-tectonic seismicity, we create a new catalog of picks by applying the volcano-trained model Volpick \cite{zhong2024deep} to the same raw data used in \cite{wilding2023magmatic}. We associate these new picks to existing picks from the \cite{wilding2023magmatic} catalog. We then recalculate locations for the events of \cite{wilding2023magmatic} using the new pick times, iteratively applying SSSTs to improve location quality as described in Sec.~\ref{sec:SSSTs}. Finally, we remove events with fewer than 10 picks to eliminate low-SNR events which are likely to suffer from greater picking error. For our study, we select events from a window encompassing the Kilauea caldera with depths between 16 km bsl and 1 km asl.  This process results in a total of 45,310 events, which yields a total of 378,074 P arrival time measurements and 280,152 S arrival time measurements. 

% \begin{figure}
%     \centering
%     \includegraphics[width=0.87\textwidth]{FiguresOld/domain.png}
%     \caption{\textbf{Source receiver geometry from a subdomain of the Island of Hawai'i.} For our real data test and realistic-geometry synthetic tests, we consider the sources and receivers in a small region of the Island of Hawai'i from \textcolor{red}{19.3$^\circ$ $-$ 19.42$^\circ$ and -155.32$^\circ$ $-$ -155.15$^\circ$} that includes the volcano of Kilauea. We use a total of 18 receivers and \textcolor{red}{17,834} events with \textcolor{red}{209,462} P arrival times and 178,692 S arrival times. These events were located using the method described in \cite{wilding2023magmatic}, but we exclude the GrowClust step. Our reconstructions have a resolution of \textcolor{red}{0.5 km}. \fix{replace the source station fig with corrected, change events, label events in legend} \fix{make this fig 1, add window}}
%     \label{fig:geo}
% \end{figure}

% \fix{We consider various initial conditions including the 1D model used to relocated the events in \cite{wilding2023magmatic} and the 3D model from \cite{lin2014three}. }
\subsection{Estimation of measurement noise model} \label{sec:noisemodel} To perform the tomography, we must define the log data likelihood as described in Eq.~\ref{eq:map} and Eq.~\ref{eq:DPI_MC}. Since we model our measurements to have additive i.i.d noise, it is sufficient to simply define a model for the measurement noise. We estimate the measurement noise by approximating a distribution over the travel time residuals from locations given by the SSSTs. We find the parameters of the residuals from P and S picks separately and find that the P residuals are best approximated as $\eta_{ij} \sim \text{Laplace}(0, b_P:=0.08)$ and S residuals are best approximated as $\eta_{ij} \sim \text{Laplace}(0, b_S:=0.14)$.
% Laplacian distribution with $b_P = 0.08$ for P picks and $b_S = 0.14$ for S picks. This gives us the following noise model $\eta_{ij} \sim \text{Laplace}(0, b_P)$ for P residuals and $\eta_{ij} \sim \text{Laplace}(0, b_S)$ for S residuals.
% \fix{Since we choose to use a Laplacian noise model assumption, this yields BLAH} 
The resulting $\log$ data likelihood is described in Sec.~\ref{sec:approach}, and we use this $\log$ data likelihood to solve Eq.~\ref{eq:map} and Eq.~\ref{eq:DPI_MC}.

\begin{figure*}
    \centering 
    \includegraphics[width=0.8\textwidth]{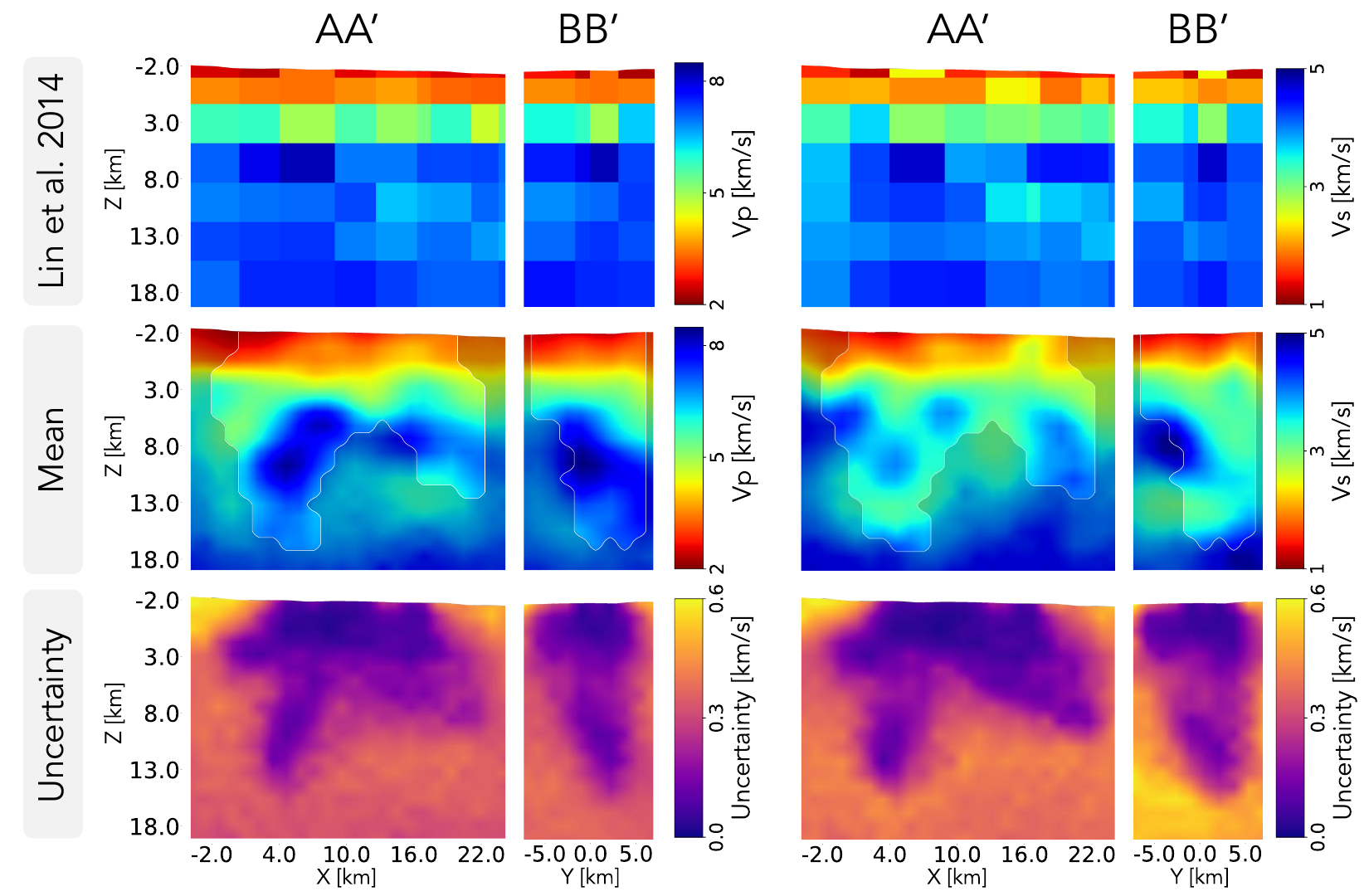}
    \caption{\textbf{Eikonal tomography of Kilauea.} We show cross-sections of results from eikonal tomography of Kilauea as described in Sec.~\ref{sec:experiments}. We show the estimated mean and uncertainty from our approach along with the results from \protect\cite{lin2014three}. The darkened part in our estimated mean is given by the resolvability index described in \ref{eq:sensitivity} where the lighter region is resolvable and darker region is not resolvable for a resolution of 4 km. Our tomography shows evidence of a high $V_P$ anomaly at around 5 to 12 km depth between 3 to 9 km along A-A'. As described in Sec.~\ref{sec:gradsmoo}, we use Gaussian smoothing of the data-likelihood's gradient to prevent overfitting and over-interpretation of heterogeneous parts of our model.}
    \label{fig:real_meas}
\end{figure*}

\begin{figure*}
    \centering 
    \includegraphics[width=.9\textwidth]{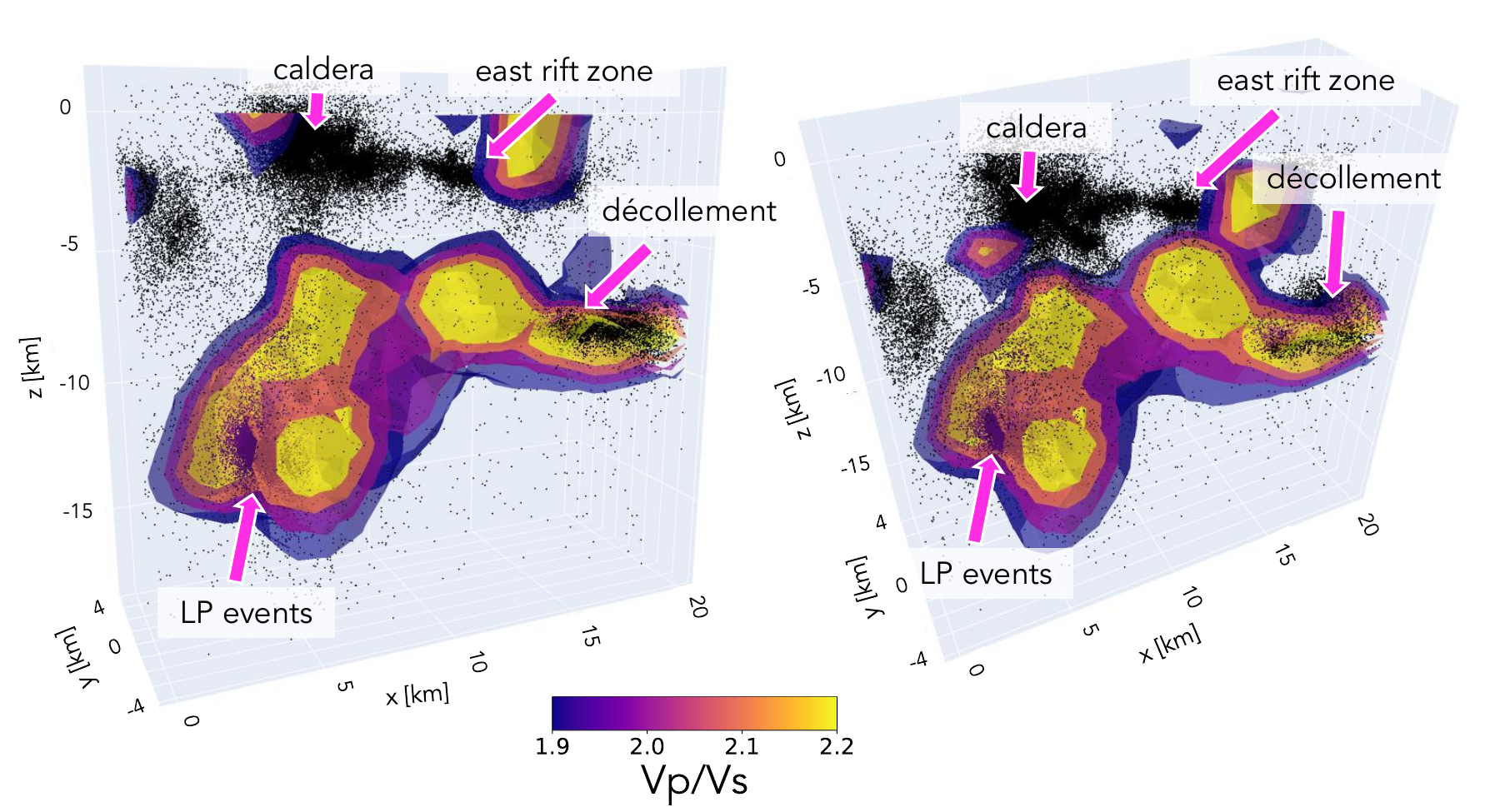}
    \caption{\textbf{High $V_P/V_S$ regions under Kilauea.} We visualize the isosurfaces of high $V_P/V_S$ regions beneath the Kilauea caldera with two views of the same volume.  This high $V_P/V_S$ structure lies between the long period events and the surface expression of the caldera. We estimate the volume of the magma column as 117 km$^3$ using the region with a $V_P/V_S \geq 2.1$, which is visualized with the orange isosurface.}
    \label{fig:isosurface}
\end{figure*}

\begin{figure*}
    \centering 
    \includegraphics[width=.8\textwidth]{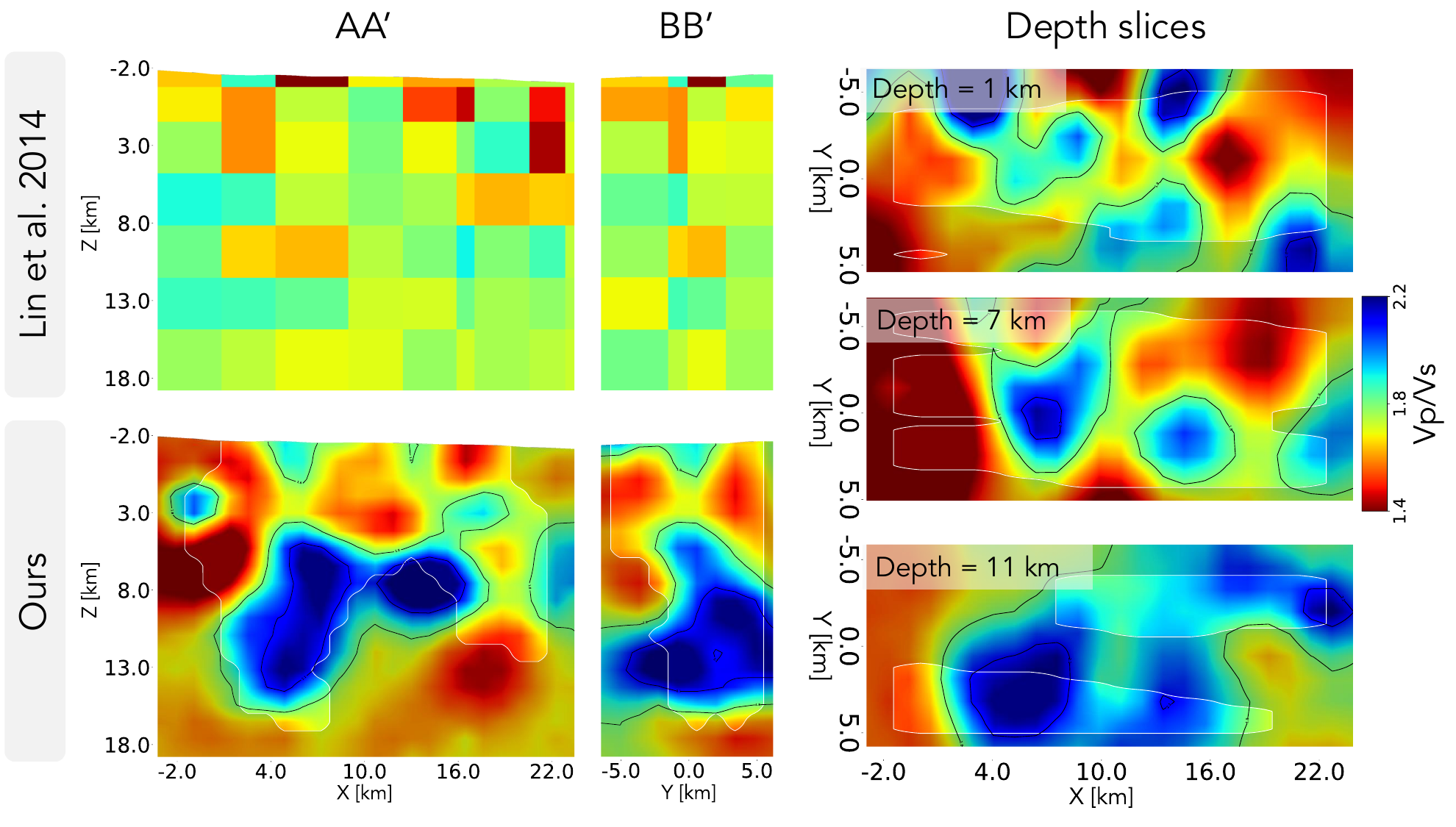}
    \caption{We show two cross sections of $V_P/V_S$ for \protect\cite{lin2014three} and the mean of our reconstructed posterior as well as additional depth slices for the mean of our reconstructed posterior. The masked region in our estimated mean is given by the resolvability index described in \ref{eq:sensitivity}. We observe a much broader range of $V_P$ and $V_S$ than in the one observed in \protect\cite{lin2014three} because we are not as dampened and invert for them separately. Our reconstruction shows evidence of a high $V_P/V_S$ chamber underlying the Kilauea caldera. Additional depth slices are included in Fig.~\ref{fig:all_depth_slices}.}
    \label{fig:ratio}
\end{figure*}

% \begin{figure*}
%     \centering 
%     \includegraphics[width=.8\textwidth]{FiguresPDFs/3D.pdf}
%     \caption{\textbf{High $V_P/V_S$ regions under Kilauea.} We visualize the isosurfaces of high $V_P/V_S$ regions beneath the Kilauea caldera with two views of the same volume.  This high $V_P/V_S$ structure lies between the long period events and the surface expression of the caldera. We estimate the volume of the magma column as 117 km$^3$ using the region with a $V_P/V_S \geq 2.1$, which is visualized with the orange isosurface.}
%     \label{fig:isosurface}
% \end{figure*}

\begin{figure}
    \centering 
    \includegraphics[width=.45\textwidth]{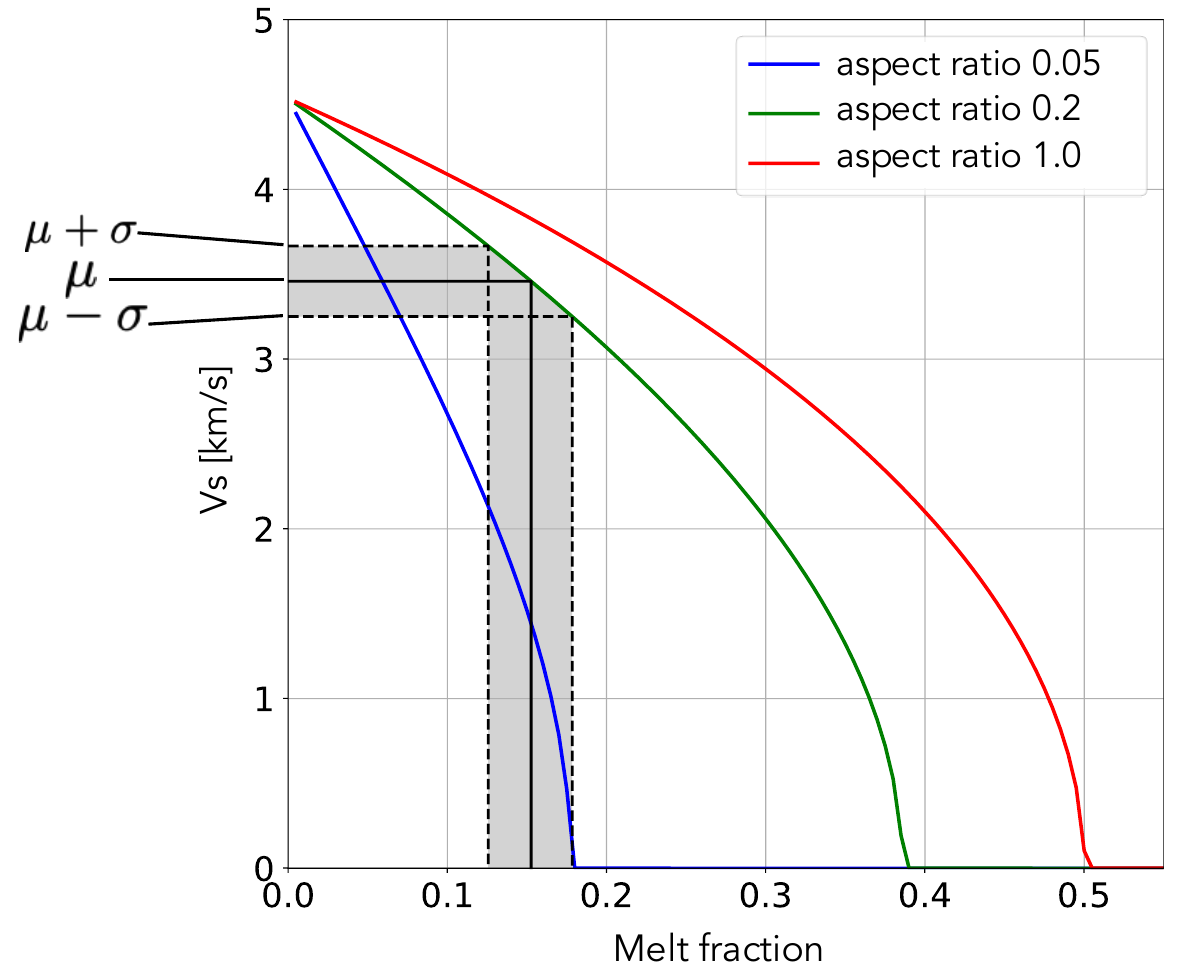}
    \caption{\textbf{Melt fraction computation with uncertainty.} We estimate the melt fraction by applying a self-consistent scheme (SCS) effective medium approach. Assuming textural equilibrium for basaltic magma with an aspect ratio of pores and cracks to be 0.2~\protect\cite{holness2006melt}, we estimate a melt fraction of 15 ± 2.7\% at 11 km depth. We compute an average $V_S$, referred to as $\mu$, and uncertainty, referred to as $\sigma$. The solid line represents the melt fraction prediction with $\mu$ and the shaded region represents 1 standard deviation $\sigma$ from $\mu$. }
    \label{fig:meltfrac}
\end{figure}

\subsection{Results}  
In Fig.~\ref{fig:real_meas} and Fig.~\ref{fig:ratio}, we show results from eikonal tomography of Kilauea using real measurements with hypocenters located using SSSTs as described in \cite{wilding2023magmatic} and Sec.~\ref{sec:SSSTs} and solve for $V_P$ and $V_S$ independently. 
% We initialize the velocity model to be the 1D model used for source localization in \cite{wilding2023magmatic}. 
% We keep the same gradient smoothing parameters as in the synthetic tests with $k=250$, $k_{max} = 500$, and $k_{stop} = 250$ as in Sec.~\ref{sec:gradsmoo_synth}.
% Our assumed noise model is $\eta_{ij} \sim \text{Laplacian}(0, b)$ where $b$ is specified in Sec.~\ref{sec:noisemodel}.
% We utilize a $\log$ normal distribution for the variational distribution to ensure that the velocity is non-negative.
Our P-velocity tomography (see Fig.~\ref{fig:real_meas}) shows evidence of a high velocity anomaly at 8 km depth under the Kilauea caldera. This feature is consistent with a previous tomography study \cite{lin2014three} which used the method of \cite{thurber1999simul} to perform tomography using locations from the Hawaiian Volcano Observatory seismic catalog [HVO]. Similar features are resolved within the S-wave model, but the different amplitudes of the imaged anomalies enable the better definition of the structures obtained from $V_P$/$V_S$ ratio, which is described in the next section. While our reconstructions are broadly similar to the features reported in \cite{lin2014three}, our more finely-spaced parametrization and dense event catalog enable a more detailed view of subcaldera velocity structure. Moreover, our method also includes the estimation of the velocity uncertainties in addition to the mean reconstruction, which can provide bounds to the accuracy of the reconstructed velocities. Synthetic checkerboard and anomaly tests (Fig.~\ref{fig:check}-\ref{fig:anomaly_partial}), which match the real data experimental setup, show that in these regions we are able to resolve features of this scale of 4 km resolution as from the checkerboard tests.

% \fix{For comparison, we also perform tomography using the raw HVO locations and the locations from just the 1D model (i.e. without SSSTs). The HVO locations provide very little information about the heterogeneities in the subsurface, but seem to imply a slight mis-estimation of the 1D velocity model in this region. The HVO locations appear to also have evidence of the same high velocity anomaly located in \fix{}, but not much other resolvable features.}

\section{Discussion}

% \subsection{Scientific findings}

% \fix{Ettore/Jack: add laymans description of the scientific result}

In this section, we present a detailed interpretation of the subsurface structures revealed by our method. We begin by comparing the imaged features with those reported in previous studies and interpreting them within the geological framework of Kīlauea. Finally, we estimate the melt fractions in the subsurface regions we identify as likely to contain partial melt.

In Fig.~\ref{fig:real_meas}, we show the mean and standard deviations of our inverted $V_P$ and $V_S$ models. In Fig.~\ref{fig:ratio}, we show $V_P$/$V_S$ for both our method and \cite{lin2014three}. We identify two regions marked by large $V_P$/$V_S$ ratios. The high $V_P$/$V_S$ region in the south flank zone around 155.2$^\circ$W and 19.32$^\circ$N and between 7-10 km depth is also found in \cite{lin2014three}.  Strikingly, we also observe high $V_P$/$V_S$ values (>2.1) along a column extending from the caldera to our lowest resolvable model depth ($\sim$15 km) that we interpret as evidence of a high-melt fraction region.  We also identify low $V_P$/$V_S$ segments at shallow depths (>0 km asl) similar to features identified by \cite{lin2014three}, who hypothesized that these anomalies reflected the presence of water-saturated pores. Additionally, we identify a high-velocity contrast beneath the rift zone at approximately 5 km depth and between 10 km to 22 km along the x axis, consistent with teleseismic receiver function analyses ~\cite{leahy2010underplating}. Below this contrast, a high $V_P$/$V_S$ anomaly centered at 8 km depth aligns with previous observations of magmatic underplating~\cite{leahy2010underplating}.

A prominent high $V_P$/$V_S$ body, located beneath Kīlauea’s crater (approximately 4 km on the x-axis in Fig.~\ref{fig:ratio}) and extending from 13 km to 4 km depth, coincides with the large magma reservoir previously inferred from seismicity and geological evidence ~\cite{ryan1987elasticity,ryan1988mechanics}. The lower extreme of this anomaly corresponds to the location of previously observed long-period earthquakes ~\cite{matoza2014high,wright2006deep,wilding2023magmatic}, suggesting that the imaged structure represents a deep magmatic pathway feeding the reservoir. To better illustrate the imaged structures, we present a 3D view of the $V_P/V_S$ isosurface in Fig.~\ref{fig:isosurface}. Along the east rift zone, deeper high $V_P$/$V_S$ bodies align with seismicity along the decollement.

To estimate the volume of the deep magma column, we calculate a $V_P/V_S$ isosurface of 2.1 and assume that volumes with $V_P/V_S$ above this value are likely melt-bearing. This process results in an estimated total volume of 117 km$^3$ for the deep magma column. To determine the melt fraction, we apply a self-consistent scheme (SCS) effective medium approach~\cite{mavko1980velocity,Paulatto2022}, assuming seismic velocities from~\cite{lin2014seismic} ($V_P$ = 7.955 km/s, $V_S$ = 4.536 km/s at 1200 K). Fig.~\ref{fig:meltfrac} displays melt fraction curves for different aspect ratios used in the SCS model. Since our tomography approach optimally fits the observed travel times while incorporating uncertainty bounds, we can quantify both the melt fraction and its associated uncertainty. Assuming textural equilibrium for basaltic magma with an aspect ratio of 0.2~\cite{holness2006melt}, we estimate a melt fraction of 15 ± 2.7\% at 11 km depth. Using the calibration curve from the SCS model, the total melt volume is estimated at 18.5 ± 4.5 km$^3$. While this value is slightly higher than previous seismic tomography estimates\cite{lin2014seismic}, the overall density of the picritic mush remains nearly unchanged at 3.2 g/cm$^3$, assuming melt and olivine densities of 2.65 g/cm$^3$ and 3.3 g/cm$^3$, respectively. As suggested in prior studies, melt likely separates from the crystal mush to ascend toward shallower depths~\cite{lin2014seismic}. 

%\fix{Jack/Ettore: add a sentence that zooms out and shows broader impacts}
Magma volumes within Kilauea’s summit reservoir have traditionally been estimated by using geophysical methods or estimating extrusive volumes; however, the magma volume estimates by these models can vary by orders of magnitude (e.g. \cite{johnson_dynamics_1992,decker_dynamics_1987,klein_patterns_1982}). Smaller estimates from derived from petrological analyses indicate that these geophysical methods may have difficulty in distinguishing between a magma chamber’s molten core and the surrounding crystal-mush, higher-temperature rock \cite{pietruszka_size_1999}. In contrast, by reducing damping and performing a high-resolution inversion, our technique allows us to identify extremely high $V_P$/$V_S$ regions beneath Kilauea that we can reliably interpret as melt-rich regions. Our results are also sensitive to melt-rich regions at depths beneath the summit reservoir and improve our understanding of the magma supply system at Kilauea. Importantly, our technique is also applicable to any volcano with sufficient seismicity. Magma supply processes are recognized as an important control on eruptive styles and timing at volcanoes \cite{caricchi_frequency_2014}; the ability to characterize the structures responsible for supply and transport may enable new insight into eruptive hazard and eruption style at other volcano complexes, as well.

%% file: sections/sec_conclusion.tex
\section{Conclusions}

% We develop an eikonal tomography method that bypasses the need for iterative earthquake relocation by leveraging SSST-based hypocenter corrections.  
Our work recovers evidence of a magmatic system underneath Kilauea at the highest resolution to date using travel-time to retrieve high-resolution models of $V_P$ and $V_S$ structures.  From our 3D tomography study, we estimate the volume of the deep (15-3 km bsl) magma column under Kilauea to be around 117 km$^3$. Our eikonal tomography method bypasses the need for iterative earthquake relocations by leveraging SSST-based hypocenter corrections. In addition, we use variational inference and stochastic optimization to estimate the point-wise uncertainty of the inverted tomographic model. Our approach enables imaging high $V_P$/$V_S$ regions in high seismicity volcanic regions, providing a pathway to better characterize complex magmatic systems and advance understanding of seismovolcanic processes.

% Although we focus on a small region for this study, future work will include studying the entire island of Hawai`i and modeling the posterior with more complex variational distributions.

%% file: sections/sec_supp.tex
\appendix
\section{Additional results}

\subsection{Full resolution tests}
\begin{figure*}
    \centering
    \includegraphics[width=\textwidth]{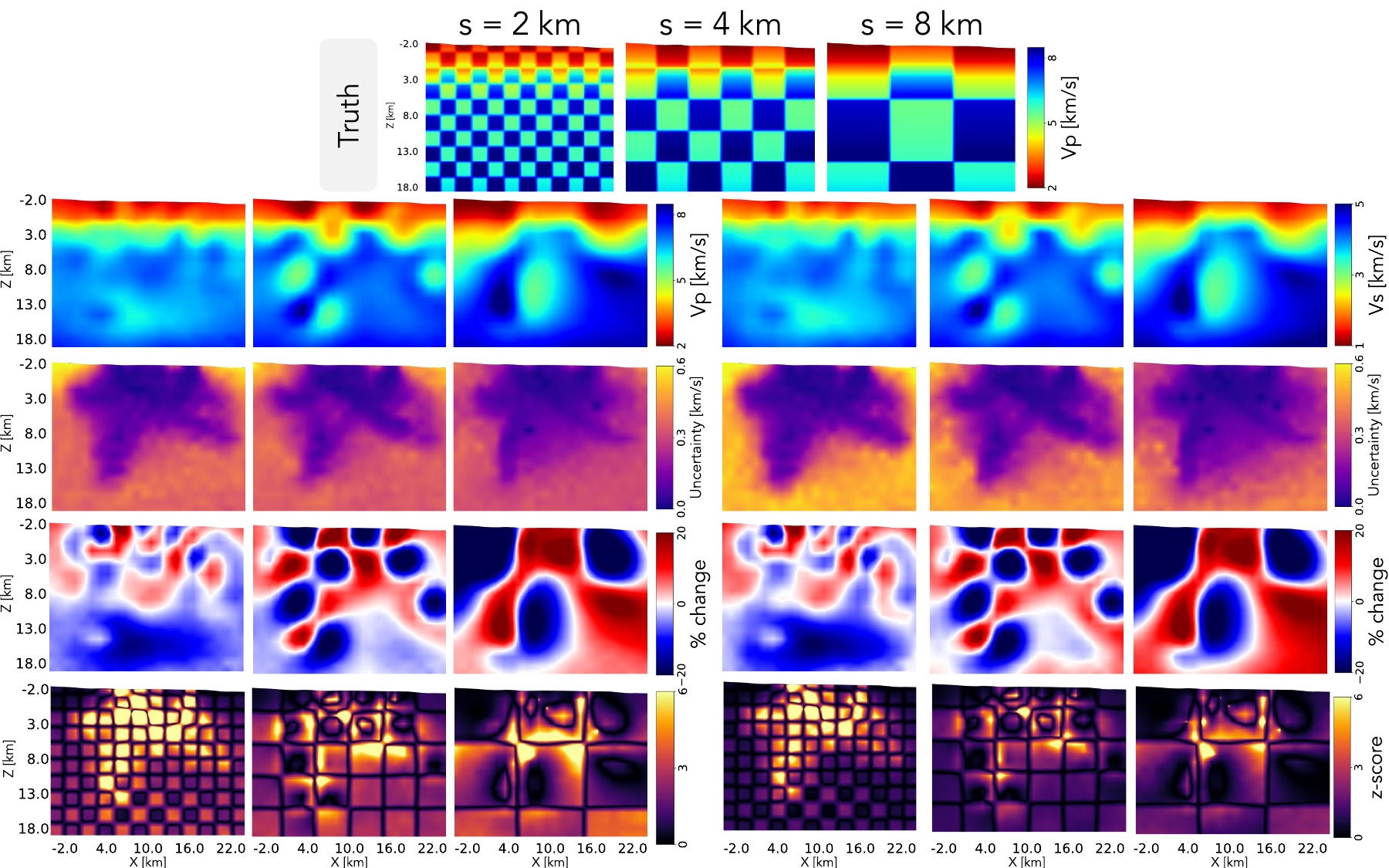}
    \caption{\textbf{Checkerboard tests}  We perform inversion on the 1D model from \protect\cite{lin2014three} modulated with various checkerboards of $25\%$ increase or decrease for both $V_P$ and $V_S$. These checkerboard patterns have a size of 2 km, 4 km, and 8 km.  We also show the reconstruction and relative change for A-A'. We are able to resolve around a resolution of 4 km in the upper half of the domain and at least a resolution of 8 km throughout the region with ray coverage for both $V_P$ and $V_S$. We include z-scores to quantify the accuracy of our uncertainty quantification. We find that the uncertainty is underestimated near the sharp interfaces of the checkerboard as well as in the high frequency regions that are below the resolvability of our method.}
    \label{fig:check_vp_vs}
\end{figure*}

In Fig.~\ref{fig:check_vp_vs}, we show the resolution tests using a checkerboard perturbation for both $V_P$ and $V_S$. We find that they have roughly the same resolvability with around $4$ km resolution everywhere that has sensitivity and up to $2$ km resolution near the surface. 

\begin{figure*}
    \centering
    \includegraphics[width=.8\textwidth]{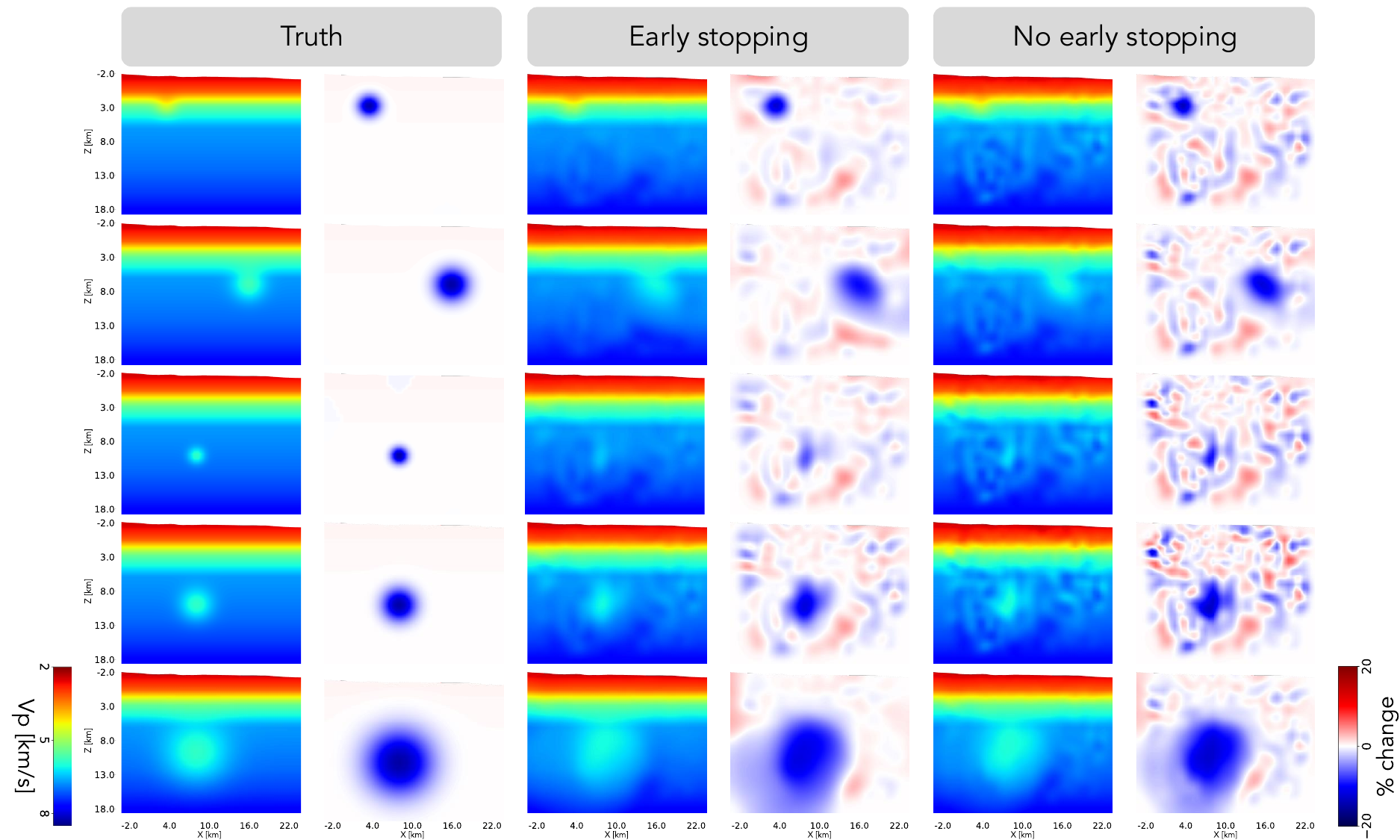}
    \caption{\textbf{Physically motivated synthetic anomaly tests.} We perform synthetic tests to explore the sensitivity of our expected source-receiver geometry to various $\sim 10\%$ low velocity anomalies in the subsurface. For various candidate anomalies, we show the following for A-A': the true velocity model used to generate the synthetic travel times, the relative change to the initial model, and the mean of the inferred posterior and relative change to the initial model for both an $k=250$ and $k=500$. Note, we plot these cross sections masking out the topography.  }
    \label{fig:anomaly}
\end{figure*}

In Fig.~\ref{fig:anomaly}, we show resolvability tests on possible anomalies.

\subsection{Optimization parameters}
\begin{figure}
    \centering
    \includegraphics[width=0.45\textwidth]{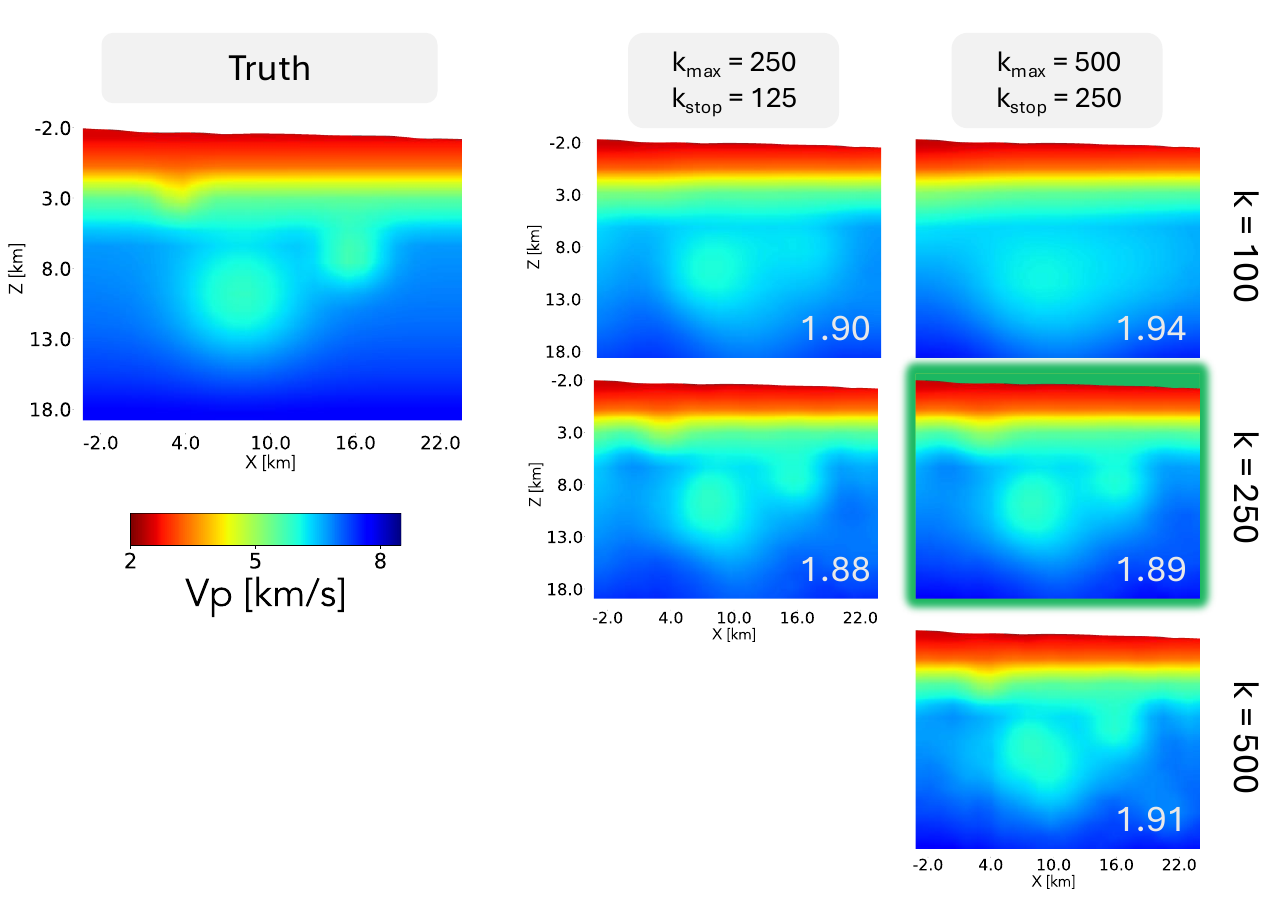}
    \caption{\textbf{Gradient smoothing.} We visualize the reconstructions along A-A' for different values of $k_{stop}$, $k_{max}$, and $k$ values for the gradient smoothing as described by Eq. ~\ref{eq:gradsmoo}. We find that the optimal set of parameters is $k_{stop}=250$, $k_{max}=500$, and $k=250$, which is highlighted in green. We include the ground truth velocity model for reference.}
    \label{fig:param_UQ}
\end{figure}

In Fig.~\ref{fig:param_UQ}, we show the optimization hyperparameters that we tested before selecting the optimal set. We use $k=250$, $k_{max} = 500$, and $k_{stop}=250$ since it fits the data while not overfitting the noise. We include the normalized data likelihood $\log p(y|V) = |y-f(V)|/b$ for each reconstruction where $b$ is the parameter of the Laplacian distribution.

\subsection{Noise model misspecification}
\begin{figure}
    \centering
    \includegraphics[width=.45\textwidth]{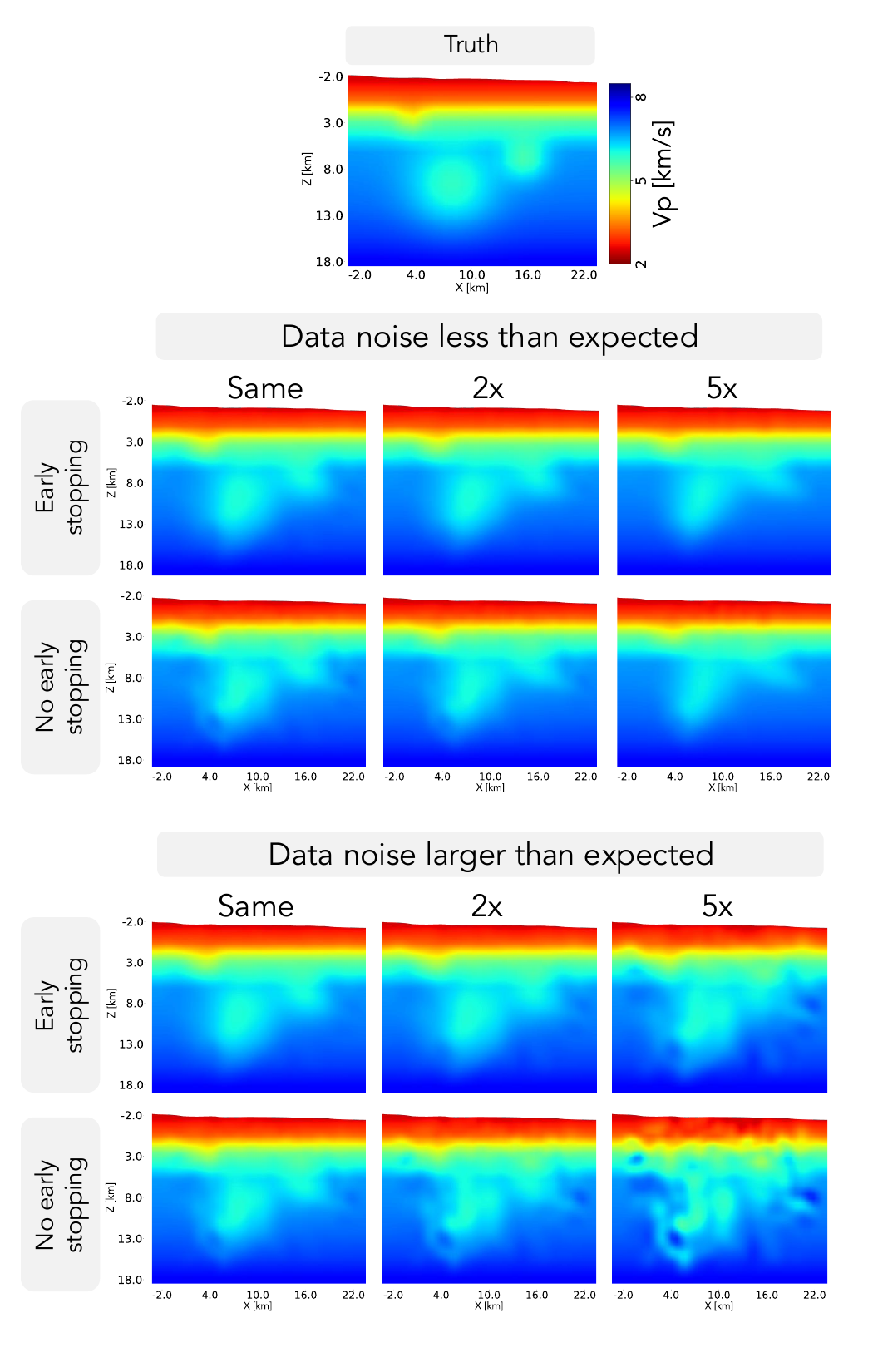}
    \caption{\textbf{Noise level mismatch.}  We consider the impact of misspecifying the noise level $b_P$ of our measurement model $\log p(y|V_P)$ when solving for $V_P$. We consider the effect of mismatch in assumed level of noise (e.g. over or under estimating). We find that when the noise is less than expected, the structures are similar but the amplitude of recovery is lower. When the noise is much higher than expected, the structures are obfuscated due to overfitting. We find it necessary to do early stopping to prevent overfitting to noise especially when the data noise is larger than expected. Truth is in Fig.~\ref{fig:param_UQ}. These are cross-sections taken along A-A'.}
    \label{fig:synth_noiselevel}
\end{figure}

In Fig.~\ref{fig:synth_noiselevel}, we show results on reconstruction $V_P$ when we have misestimated the noise by a factor of 2 and 5. In all settings, we are able to resolve the anomalies, but when the noise is much higher than expected, there is substantial overfitting to noise. We find it necessary to do early stopping to prevent overfitting to noise, especially when the data noise is larger than expected. 

\begin{figure}
    \centering
    \includegraphics[width=.45\textwidth]{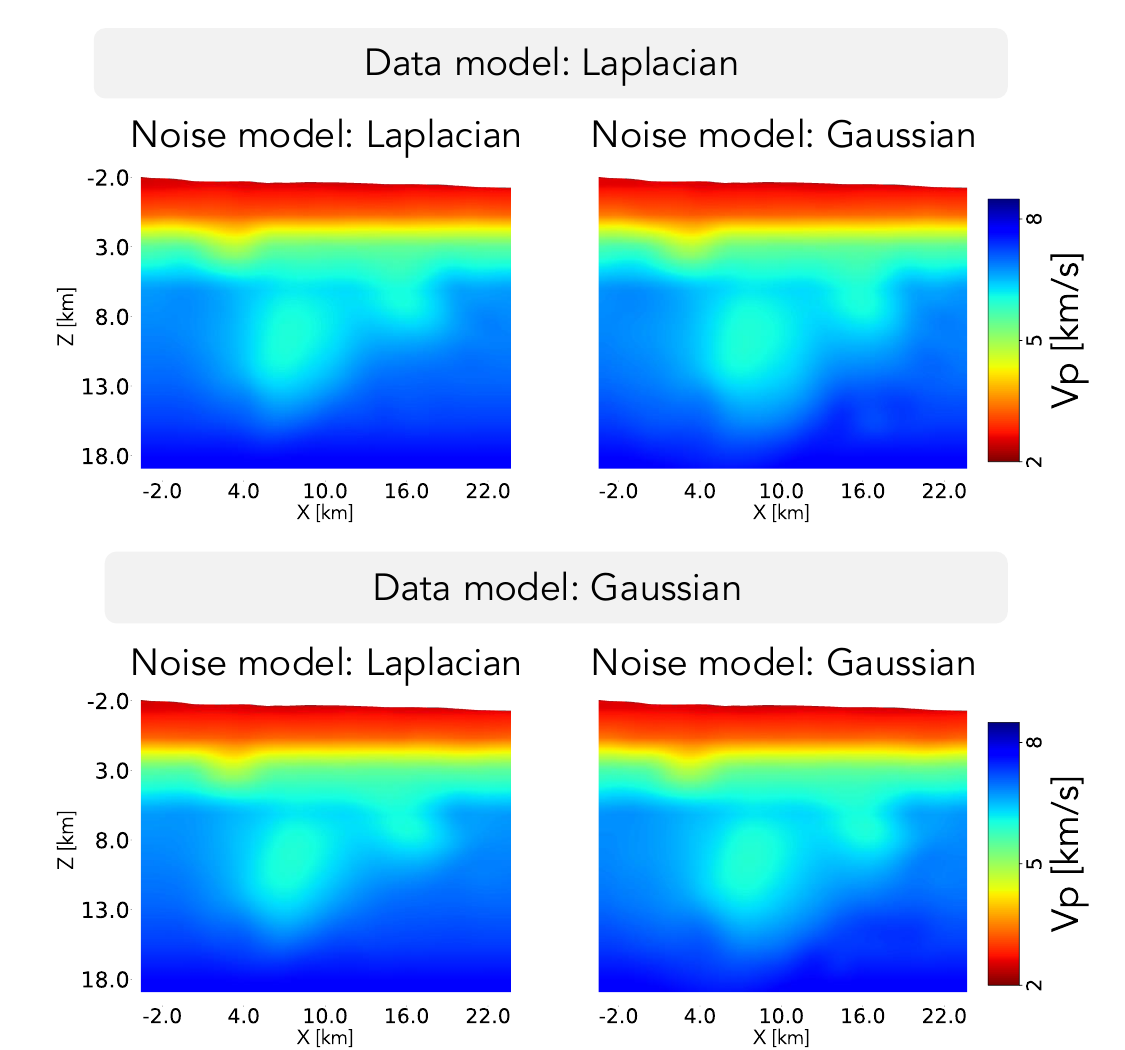}
    \caption{\textbf{Noise model mismatch}  We consider the impact of misspecifying the noise model $\eta$ and/or measurement model $\log p(y|x)$. We consider the effect of mismatch in the assumed noise model distribution (e.g. Gaussian vs. Laplacian noise). For the Gaussian distribution, we assume a standard deviation $\sigma_P = .11$ and $\sigma_S=0.31$. For the Laplacian distribution, we assume $b_P=0.08$ and $b_S = 0.15$. The noise model and data-fit model highlighted in green is the one that we expect our real data to match. Truth is in Fig.~\ref{fig:param_UQ}.}
    \label{fig:synth_mismatch}
\end{figure}

We also consider a mismatch in the noise model itself and consider Laplacian and Gaussian distributions as shown in Fig.~\ref{fig:synth_mismatch}. 

\begin{figure*}
    \centering
    \includegraphics[width=.8\textwidth]{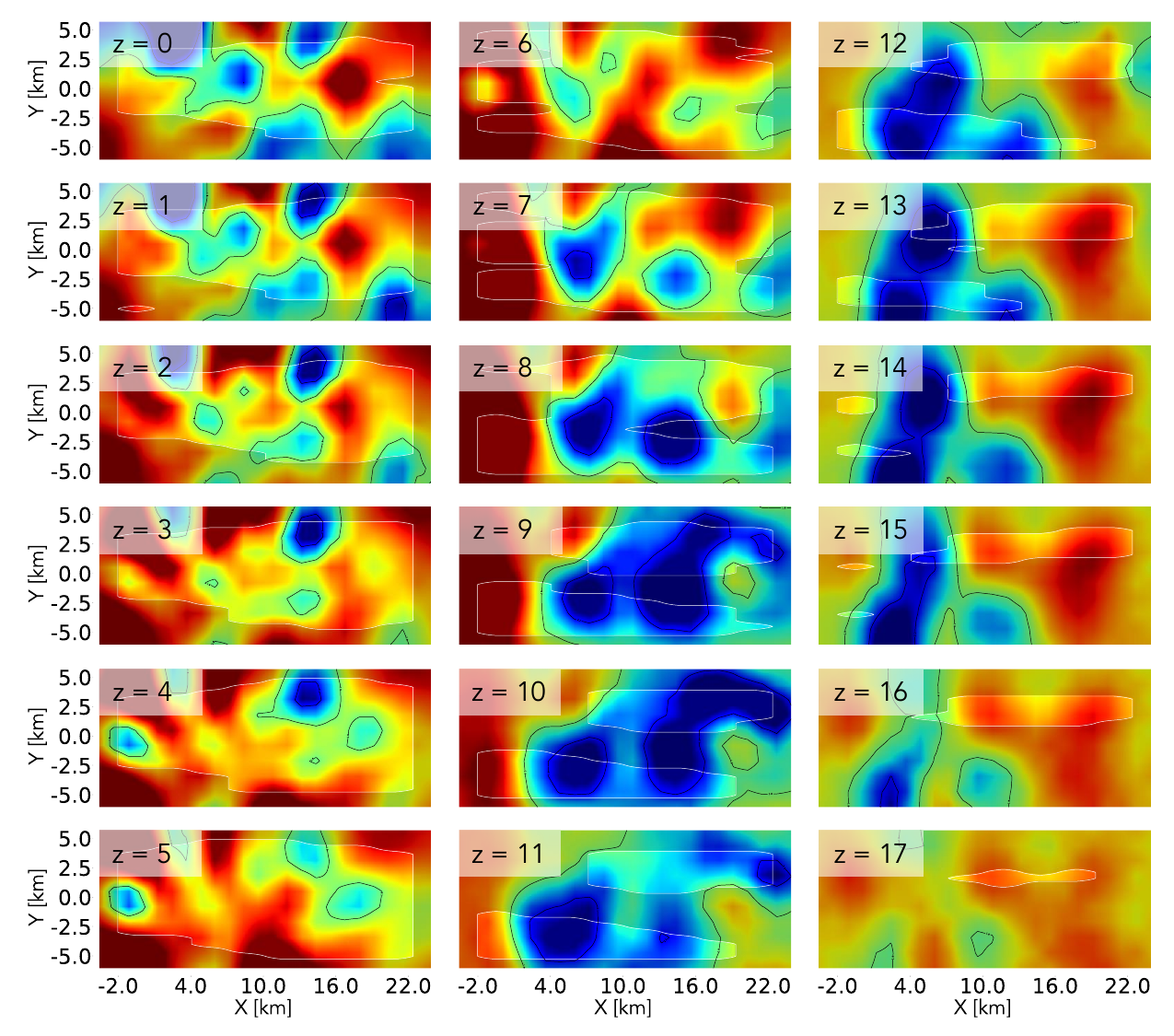}
     \caption{\textbf{All depth slices of $V_P/V_S$.} Here we show all depth slices of $V_P/V_S$ from $z=0$ km to $z=17$ km. Refer to the color map in Fig.~\ref{fig:ratio} or Fig.~\ref{fig:sssts}. The lighter region indicates resolvability with a checkerboard of $4$ km.}
    \label{fig:all_depth_slices}
\end{figure*}

\begin{figure*}
    \centering
    \includegraphics[width=.7\textwidth]{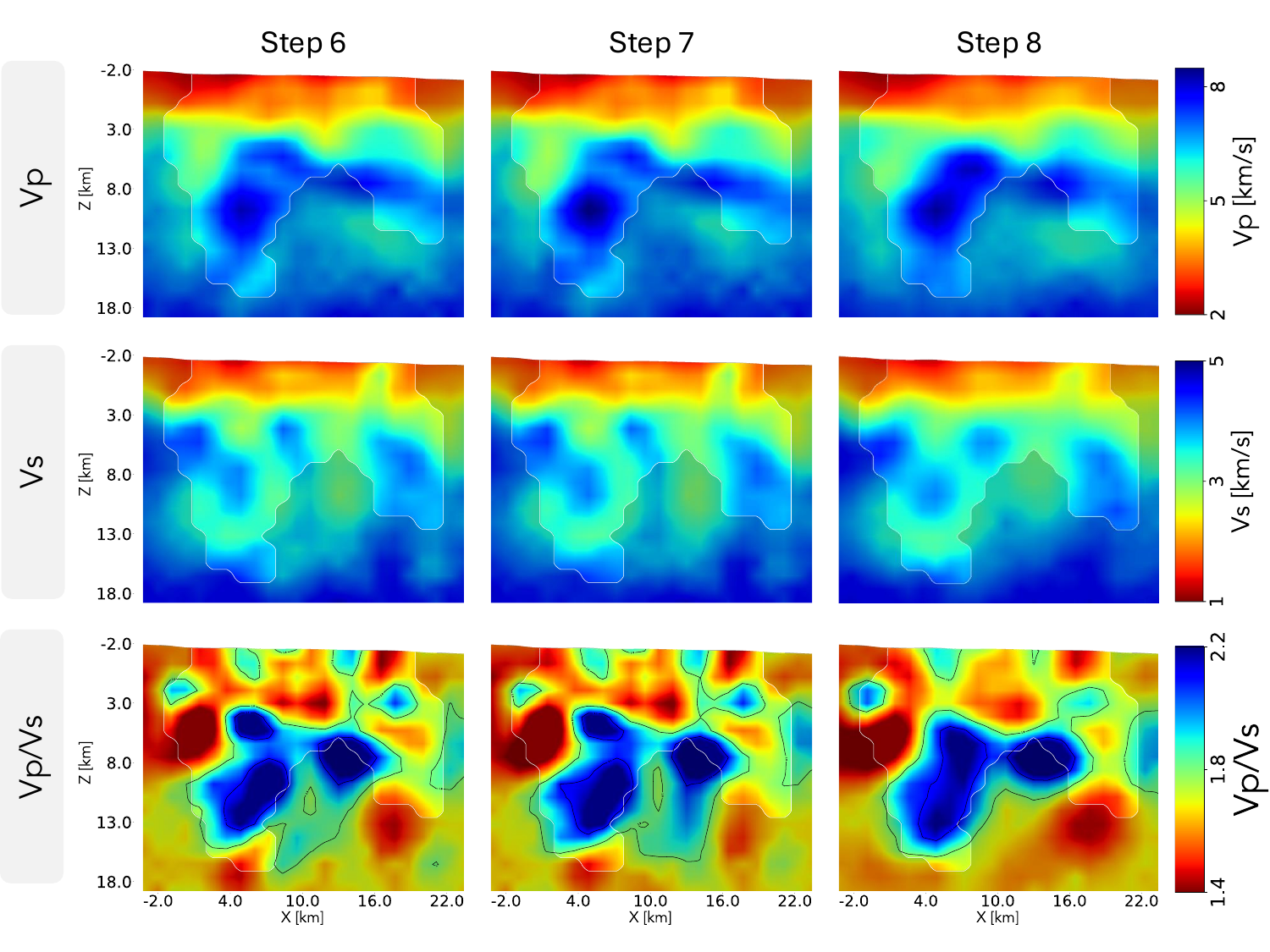}
     \caption{\textbf{Reconstructions at various SSST steps.} Here we show reconstructions at various steps in the SSST localization procedure along the A-A' cross section. We find that the recovered structures are stable across these different steps.}
    \label{fig:sssts}
\end{figure*}

\section{Theoretical perspective of gradient blurring for linear inverse problems}
% \begin{itemize}
% \item Implicit and explicit gradient blurring:
% \item Implicit: through augmentation (normalizing flows blur the gradient with the quantization error, diffusion blurs the gradient with the SDE)
% \item Implicit: regularizers can change the gradient to be blurred
% \item What if we actually blur the spatial gradient instead? how does this relate to the implicit regularization in NeRF?
% \end{itemize}

% \subsection{Linear inverse problems}
\subsection{Gradient smoothing for linear inverse problems}\label{sec:linear_gradsmoo}
We consider solving a linear inverse problem where $A\in \mathbb{R}^{n^2 \times  n^2}$, $\mathbf{v} \in \mathbb{R}^{n^2 \times 1}$ is our target, $y \in \mathbb{R}^{n^2 \times 1}$ is our measurements, and $\mathcal{R}(\mathbf{v})$ is our regularizer by solving the following optimization problem:
\begin{align}
    \hat{\mathbf{v}} &= \arg\min_{\mathbf{v}} \mathcal{L}\nonumber\\
    &= \arg\min_{\mathbf{v}} ||y - A\mathbf{v}||_2^2 + \mathcal{R}(\mathbf{v}).
\end{align}
For simplicity, we assume that the regularizer is 0:
\begin{align}
    \hat{\mathbf{v}} = \arg\min_{\mathbf{v}} ||y - A\mathbf{v}||_2^2 .
\end{align}
If we solve this using gradient descent, then the gradient is defined by:
\begin{align}
    \frac{\partial\mathcal{L}}{\partial\mathbf{v}} = -A^T(y-A\mathbf{v}) .   
\end{align}

We consider adding an isotropic spatial Gaussian blur $G$ to the data-likelihood that can be decomposed as $G=G_{\text{1D}}G_{\text{1D}}^T \in \mathcal{R}^{n^2 \times n^2}$. G is self-adjoint. Applying $G$ to the gradient gives us:
\begin{align}
    \frac{\partial\mathcal{L}}{\partial\mathbf{v}} &= G\left[-A^T(y-A\mathbf{v})\right]\nonumber\\   
    &=-GA^T(y-A\mathbf{v}) .   
\end{align}
Essentially, this is saying that the update from the data should be blurred because of our inductive biases (i.e. implicit smoothness prior on the gradient, potential artifacts of the gradient). Essentially, what this is doing is that it attenuates any high frequency content from the gradient of the data fit. Thus, this blurred gradient only updates the low frequencies of $\mathbf{v}$, forcing the high frequencies of the initialization to be constant. In this scheme, it is important to make sure that the frequency content of $\mathbf{v}$ matches the frequency content of the blurring operator $G$ or else the reconstructed $\mathbf{v}$ will retain these high frequency features.

% This forces newly updated $v$ to contain the same high frequency content as the initialization, but allows for flexibility to correct low frequency errors.
\subsection{Hierarchical approaches for linear inverse problems}\label{sec:linear_hierarchy}
If instead we want to improve the reconstruction by employing a hierarchical approach, then this would result in the following optimization where $G\mathbf{v}$ represents the inversion at a specific level in the hierarchical approach:
\begin{align}
    \hat{\mathbf{v}} &= \arg\min_{\mathbf{v}} \mathcal{L} \nonumber\\
    &= \arg\min_{\mathbf{v}} ||y - AG\mathbf{v}||_2^2
\end{align}
\begin{align}
    \frac{\partial\mathcal{L}}{\partial\mathbf{v}} = -G^TA^T(y-AG\mathbf{v}). 
\end{align}
Note that G is self-adjoint (i.e. $G^T = G$). Here, the gradient appears to be different than that in Sec.\ref{sec:linear_gradsmoo}, but if $G\mathbf{v} = \mathbf{v}$, then they are equivalent. This only occurs when $\mathbf{v}$ is band-limited relative to the filter $G$.

\begin{figure}
    \centering 
    \includegraphics[width=0.45\textwidth]{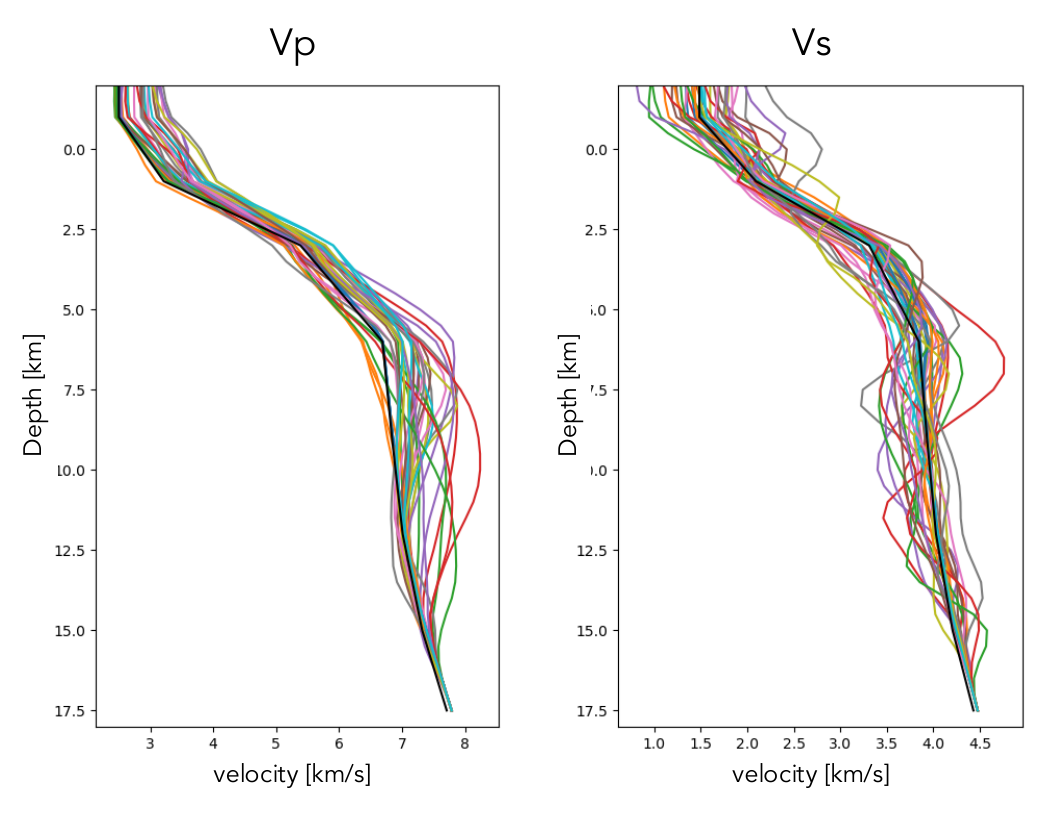}
    \caption{\textbf{1D profiles}. Various depth slices across different lat-lon locations. The black line is the 1D model from \protect\cite{wright2006deep} that we use for our initial model.  }
    \label{fig:1D}
\end{figure}